\documentclass[10pt,twocolumn,amsmath,amssymb,aps,prb,showpacs]{revtex4-2}
\usepackage{mathrsfs}
\usepackage{graphicx}
\usepackage{dcolumn}
\usepackage{bm}
\usepackage{paralist}
\usepackage{float}
\usepackage{mdframed}
\usepackage{booktabs}
\usepackage{mathtools} 
\usepackage[normalem]{ulem}
\usepackage[colorlinks,linkcolor=blue,anchorcolor=blue,citecolor=green]{hyperref} 
\usepackage{cleveref}
\usepackage{url}
\usepackage{color}
\usepackage{verbatim}
\usepackage{comment}
\usepackage{multirow}

\newcommand{\e}{\mathrm{e}}

\newcommand{\mi}{\mathrm i}
\newcommand{\dd}{\mathrm d}
\renewcommand{\exp}{\mathrm{exp}}
\newcommand{\no}{\nonumber}
\newcommand{\R}{\operatorname{Re}}
\newcommand{\I}{\operatorname{Im}}

\begin{document}
	
\title{Quantum critical properties of non-Hermitian XY models with magnetic field}

\author{Jia-Jia Luo} 
\email{jjluo@physik.rwth-aachen.de}
\affiliation{Institut für Theorie der Statistischen Physik, RWTH Aachen University, 52056 Aachen, Germany}

\author{Volker Meden}
\email{meden@physik.rwth-aachen.de}
\affiliation{Institut für Theorie der Statistischen Physik, RWTH Aachen University, 52056 Aachen, Germany}
\date{\today}

\begin{abstract}
	The characterization of the quantum critical properties of genuine non-Hermitian many-body systems remains ambiguous as neither the state considered nor the definition of expectation values is unique. In this work, we investigate the quantum critical properties of two models of non-Hermitian XY spin chains with magnetic field. Using exact solutions, we systematically investigate the parameter dependence of the energy, the magnetization as well as the long-distance asymptotic behavior of static correlation functions. We compute expectation values within the standard formalism of quantum mechanics as well as within biorthogonal quantum mechanics and take two different states which one might reasonably consider to be the analog of the ground state of a Hermitian model. The critical properties, including such fundamental characteristics as the phase diagram, depend on both the formalism used as well as the state considered. We provide arguments in favor of the use of standard quantum mechanics. Which state to be taken in computations, depends on the (hypothetical) experimental preparation of the system.   
\end{abstract}

\maketitle

\section{Introduction}	

Quantum phases and the corresponding phase  transitions are two of the most fascinating manifestations of emergent many-body phenomena \cite{Sachdev2011}. The physics becomes even richer when non-Hermitian Hamiltonians are studied \cite{Ashida2020,Rotter2009,Benderbook,Meden2023}. Non-Hermitian Hamiltonians might appear when focusing on system degrees of freedom of a (open) quantum system coupled to an environment. However, as we will elaborate on in Sect.~\ref{sec:open_sys}, one has to be rather precise in defining open system setups which can show genuine non-Hermitian physics; they require constant observation. Even when focusing on setups which obey this requirement, two rather fundamental questions were not answered satisfactorily. 

In non-Hermitian quantum theory two formalism were used in the past when defining observables and computing expectation values. One is standard quantum mechanics and the other is so-called biorthogonal quantum mechanics \cite{Brody2014} (or variants of this \cite{Benderbook}). This issue concerns quantum systems in general \cite{Meden2023} and quantum critical properties of many-body systems in particular (see below). Both formalism were used when investigating the latter, but, up to rare cases, not for the same model. We here use both formalism for the same models and systematically show that the results partly differ. We provide strong evidence that for (at least in principle) experimentally realizable genuine non-Hermitian quantum systems standard quantum mechanics should be used. This complements earlier related work on quantum many-body systems not focusing on critical properties \cite{Meden2023}. 

When investigating temperature $T=0$ quantum critical properties of Hermitian many-body systems the state to consider is the ground state. The eigenenergies of generic non-Hermitian Hamiltonians are complex. One thus loses the concept of a (many-body) state with lowest energy (the ground state). This raises the very fundamental question which state to consider when investigating the critical properties of non-Hermitian systems. In many papers on non-Hermitian quantum systems this question is not even addressed. We carefully investigate this and focus on two states which can reasonably be considered as experimentally accessible extensions of the ground state to a non-Hermitian system. We show that the critical properties detected in observables and correlation functions depend on the state, even on a very fundamental level---the phase diagrams might differ. Which type of critical properties are observable (at least in principle) in an experimental setup will depend on the preparation of the system.   

We investigate the research questions raised above focusing on one-dimensional non-Hermitian XY spin-1/2 chains (for Hermitian XY models see \cite{Auerbach2016,Sachdev2011}). For such we are able to derive closed analytical expressions for the energy density, the magnetization, and static correlation functions which provide access to the critical properties. These expressions allow for a numerical evaluation up to arbitrary precision and frequently also to analytical insights on, e.g., the asymptotic behavior of correlation functions. 

A variety of non-Hermitian XY models and close variants were studied in the past, mostly relying on numerical evaluations of the closed analytical expressions; for references, see below. However, partly for the reasons mentioned above (formalism used, state considered) and partly because of an improper interpretation of the numerical data, the status of the field is far from being satisfying. Complementing numerical data by analytical insights we here provide a comprehensive picture of the critical properties of two specific non-Hermitian XY models. Our methods can be used for other such models as well.       

\subsection{The Hermitian XY model with magnetic field}
\label{sec_Herm}
One of the prototypical models of quantum statistical mechanics to study temperature $T=0$ phases and the associated quantum phase transitions is the one-dimensional (1D) anisotropic XY spin-1/2 chain with a magnetic field along the $z$-direction. The general form of the Hamiltonian with nearest-neighbor spin coupling is
\begin{equation}
H = -\frac{J}{2}\sum_{l=1}^N \left( \frac{1+\gamma}{2} \, \sigma_l^x \sigma_{l+1}^x 
+ \frac{1-\gamma}{2} \, \sigma_l^y \sigma_{l+1}^y \right)
- \frac{\lambda}{2} \sum_{l=1}^N \sigma_l^z, \label{Ham}
\end{equation}
where the $\sigma_l^{\alpha} (\alpha=x,y,z)$ are the Pauli matrices at site $l$ on a 1D lattice of $N$ sites. We here assume periodic boundary conditions with $\sigma_{l+N}^{\alpha}=\sigma_l^{\alpha}$. The model parameters are the exchange amplitude $J$, the anisotropy $\gamma$, and the magnetic field $\lambda$. In the following we set $J=1$ and measure energies in units of $J$. 

The anisotropic XY model with vanishing magnetic field was analytically solved by Lieb, Schultz and Mattis employing a Jordan-Wigner transformation \cite{Lieb1961}. Building on this the  statistical mechanics properties of the full model were systematically studied in a series of papers by Barouch and McCoy \cite{Barouch1970a,Barouch1970b,Barouch1970c,Barouch1970d}. 

\begin{figure}[t] 
	\begin{center} 
		\includegraphics[width=0.9\linewidth]{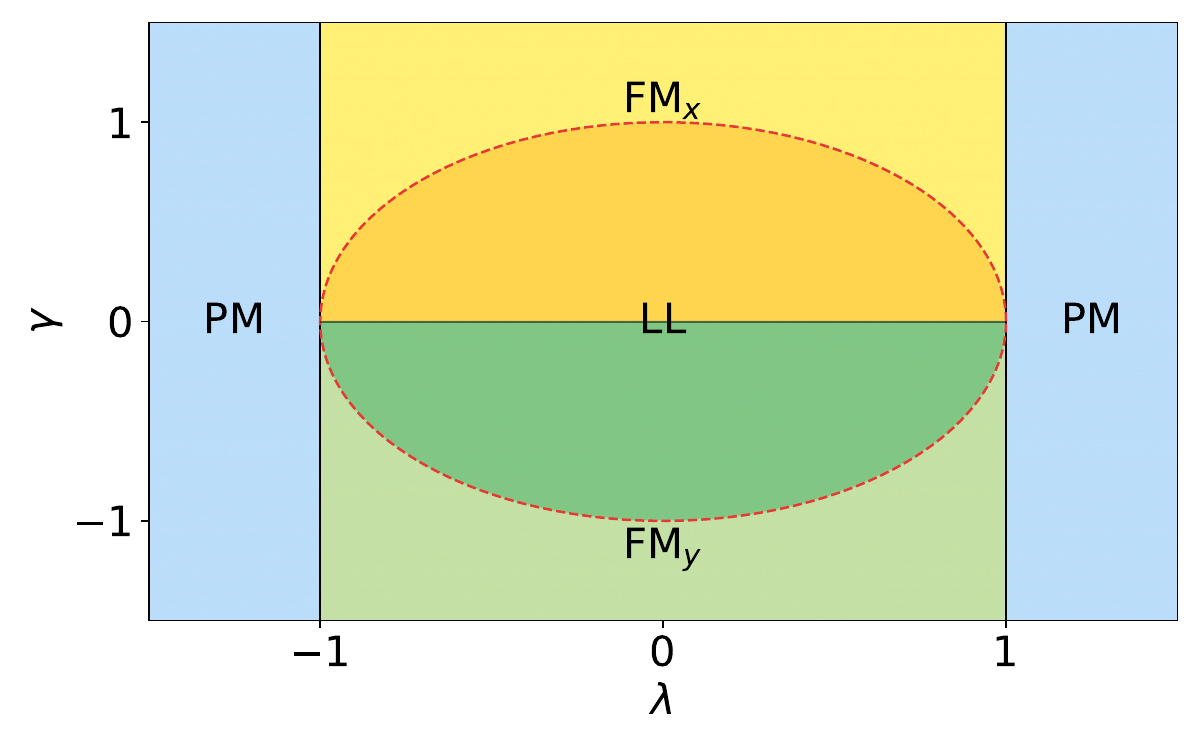}
	\end{center}
	\vspace{-14pt}  
	\caption{The ground state phase diagram of the Hermitian XY model with anisotropy $\gamma$ and magnetic field $\lambda$. The black lines indicate the phase transition lines at $|\lambda|=1$ and $\gamma=0$. The red dashed line, defined by $\lambda^2+\gamma^2=1$, marks the boundary between oscillatory and non-oscillatory behavior of the correlation function.}         
	\label{pd_her}
\end{figure}

To be self-contained and for later reference we here summarize the (Hermitian) models critical properties. The $T=0$ phase diagram of this XY model in the $(\lambda, \gamma)$-plane is shown in Fig.~\ref{pd_her}. For $|\lambda| > 1$, the system is in a gaped paramagnetic (PM) phase. At large distances $r$ between the two spins the $x$ and $y$ correlation functions $C_r^x$ and $C_r^y$, defined by the ground state expectation value 
\begin{align}
C^{\alpha}_r=\langle\sigma_l^{\alpha}\sigma_{l+r}^{\alpha}\rangle,
\label{correl_fun}
\end{align}
decay exponentially to zero. The correlation length $\xi$ is given by $\xi^{-1} = |\ln |z_*||$ with \cite{Barouch1970b,Franchini2017}
\begin{equation}
z_*=\frac{\lambda+\operatorname{sgn}(\lambda) \sqrt{\lambda^2+\gamma^2-1}}{1 + \gamma}.
\label{her_z}
\end{equation}

The two phase transition lines at $|\lambda|=1$ are from the Ising universality class and at these the correlation functions $C_r^{x/y}$ asymptotically decay as  power laws; for details see Table \ref{tab_her}. 

In the region $|\lambda|<1$, the system is gapped and exhibits long-range order if $\gamma \neq 0$. For $\gamma >0$ the order is ferromagnetic in $x$-direction (denoted as FM$_x$). The $x$ correlation function approaches a constant at large $r$ while $C_r^{y}$ goes to zero exponentially. For $\gamma<0$ the system is ferromagnetic in $y$-direction (FM$_y$) with interchanged asymptotics of $C_r^{x/y}$; see Fig.~\ref{pd_her} and Table \ref{tab_her}. 

These two ferromagnetically ordered phases are separated by the gapless Luttinger Liquid (LL) phase at $\gamma = 0$, corresponding to the anisotropic transition. In this $C_r^{x/y}$ decay to zero as $r^{-1/2}$ (quasi-long-range order). 

\begin{table}[htbp]
	\centering
	\begin{tabular}{cccc}
		\toprule\toprule
		Phase & Condition & $C_r^x$ & $C_r^y$ \\
		\midrule
		LL & $\gamma = 0, |\lambda| < 1$ & $r^{-1/2}$ & $r^{-1/2}$ \\	\midrule
		\multirow{3}{*}{FM} & $\gamma^2 + \lambda^2 < 1$ & const. & $e^{-2r/\xi}$ \\
		& $\gamma^2 + \lambda^2 = 1$ & $\gamma/2(1+\gamma)$ & $0$ \\
		& $\gamma^2 + \lambda^2 > 1, |\lambda| < 1$ & const. & $e^{-2r/\xi}$ \\	\midrule
		FM-PM & $|\lambda| = 1$ & $r^{-1/4}$ & $r^{-9/4}$ \\ \midrule	
		PM & $|\lambda| > 1$ & $e^{-r/\xi}$ & $e^{-r/\xi}$ \\
		\bottomrule\bottomrule
	\end{tabular}
	\caption{Asymptotic behavior of the correlation functions $C_r^x$ and $C_r^y$ in the Hermitian XY model with magnetic field for anisotropy $\gamma \geq 0$. The results for $\gamma<0$ can be obtained by interchanging $x$ and $y$.}\label{tab_her}
\end{table}

On a more refined level and in the FM phases one can further classify the asymptotic behavior of the correlation functions. The red dashed curve in Fig.~\ref{pd_her}, given by $\lambda^2+\gamma^2=1$, marks the boundary between oscillatory and non-oscillatory correlation functions, with oscillations occurring inside the ellipse (dark color).

\subsection{Critical properties of open systems}
\label{sec:open_sys}

The critical properties of an open quantum system coupled to an environment can be studied from different perspectives \cite{Breuer2002,Daley2014,Ashida2020,Rotter2009,Meden2023}. In a first step to investigate the physics of such setups, one usually aims at reducing the degrees of freedom to be treated to those of the system. To achieve this different techniques are available, often relying on certain approximations. Their validity must be considered from case to case.   

Employing the Green function formalism the environment can be projected out using Feshbach projection \cite{Ashida2020,Rotter2009,Taylor2006}. This leads to a self-energy $\Sigma$ complementing the Hamiltonian $H_{\rm s}$ containing the system degrees of freedom. Feshbach projection is often used for the coupled system being in total equilibrium (within Matsubara formalism), in a non-equilibrium steady state as well as in a transient non-equilibrium situation (both in Keldysh formalism). To further illustrate this approach, let us focus on the equilibrium case. In general the self-energy will be a function of the Matsubara frequency. If one now approximates $\Sigma(\mi \omega)$ to be a constant $\Sigma$ one can interpret $H_{\rm s} + \Sigma$ as an effective Hamiltonian, which, however, will generically be non-Hermitian. One can then use the language of non-Hermitian quantum mechanics when, e.g., investigating spectral properties \cite{Meden2023,Rotter2009,Brody2014,Ashida2020} (see also below). However, as long as the time-evolution is not performed with the non-Hermitian Hamiltonian, what one obtains this way is merely an approximate reformulation of the problem which could as well be treated sticking to the standard Green function formalism of Hermitian quantum mechanics. While using the language of non-Hermitian quantum mechanics might turn out to be advantageous to investigate the physics of such setups (for a recent example see, e.g., \cite{Shen2024}), they are not genuinely non-Hermitian (see below) and accordingly do not show interesting non-Hermitian physics. We here do not consider such setups any further. 

A different situation can be reached if one considers a quantum master equation for the time evolution of the systems reduced density matrix $\rho_{\mathrm{s}}(t)$ \cite{Breuer2002,Daley2014}. Assuming a weak system-environment coupling and a clear separation of the system and environment times scales as well as taking the Markov approximation the Lindblad master equation 
\begin{align}
\mi \partial_t \rho_{\rm s}(t)  = &  \left[H_{\rm s},\rho_{\rm s}(t)\right] \label{eq:lindblad} \\ &  - \frac{\mi}{2} \! \sum_{l} \! \kappa_l  \! \left\{ L_l^\dag L_l \rho_{\rm s}(t) + \rho_{\rm s}(t) L_l^\dag L_l - 2 L_l \rho_{\rm s}(t) L_l^\dag \right\}   \nonumber
\end{align}
holds. Here the $L_l$ are the Lindblad jump operators describing loss and gain processes with rates $\kappa_l$. Often, the jump operators and the rates can only be constructed based on phenomenological reasoning. In fact, implementing a desired set of jump operators and rates may require continuous external control of the system-environment coupling. The three approximations leading to Eq.~\eqref{eq:lindblad} are often met in AMO systems \cite{Daley2014,Ashida2020}. A typical example is a few-level atom coupled to a quantized light field.
  
The full Lindblad master equation for spin models was investigated in several studies \cite{Joshi2013,Prosen2008,Feig2013,Znidaric2010,Monthus2017,Maghrebi2016}. In particular, the density matrix of the steady state was computed and its critical properties were studied \cite{Joshi2013,Prosen2008,Znidaric2010,Monthus2017,Maghrebi2016}. The notion of non-Hermiticity does not appear in such studies. 

Non-Hermitian physics comes into play if one rewrites Eq.~\eqref{eq:lindblad} as
\begin{equation}
\mi \partial_{t} \rho_{\mathrm{s}}(t) = H_{\mathrm{eff}} \rho_{\mathrm{s}}(t) - \rho_{\mathrm{s}}(t) H_{\mathrm{eff}}^{\dagger} - \frac{\mi}{2} \sum_{l} \kappa_l L_{l} \rho_{\mathrm{s}}(t) L_{l}^{\dagger},\label{master}
\end{equation}
with the corresponding effective non-Hermitian Hamiltonian $H_{\rm eff}$ given by
\begin{equation}
H_{\mathrm{eff}} = H_{\mathrm{s}} - \frac{\mi}{2} \sum_{l} \kappa_l L_{l}^{\dagger} L_{l}.\label{lind}
\end{equation}
If one neglects the so-called recycling term given by the last addend in Eq.~\eqref{master}, the dynamics of the reduced systems density matrix is governed solely by the non-Hermitian effective Hamiltonian \eqref{lind}. Neglecting the recycling term is justified for times in which no jump process occurred. 

The protocol in which $H_{\rm eff}$ determines the dynamics is thus the following: The system is under permanent observation. Only instances in which no jump occurred are kept; one often refers to this as post selection \cite{Zhu2011}. Crucially the post selection (observation/measurement) affects the dynamics and the time evolution of the reduced density matrix is given by 
\begin{equation}
\mi \partial_{t} \rho_{\mathrm{s}}(t) = H_{\mathrm{eff}} \rho_{\mathrm{s}}(t) - \rho_{\mathrm{s}}(t) H_{\mathrm{eff}}^{\dagger} . \label{master_nojump}
\end{equation}
As $H_{\mathrm{eff}} \neq H_{\mathrm{eff}}^{\dagger}$ this equation cannot be rewritten as a commutator as it is the case in the standard von Neumann equation. Remarkably, the time evolution is determined by the non-Hermitian Hamiltonian $H_{\rm eff}$, even though no jump---no loss or gain---took place. 

For a pure state of the system $\rho_{\mathrm{s}}(t) = \left| \psi(t) \right> \left< \psi(t) \right|$ the dynamics is given by the standard Schr\"odinger equation with non-Hermitian Hamiltonian
\begin{equation}
\mi \partial_{t} \left| \psi(t) \right>  = H_{\mathrm{eff}}   \left| \psi(t) \right> . \label{non_herm_SG}
\end{equation}

It is this protocol, involving a continuous measurement, combined with the engineering possibly required to realize certain jump operators, which leads to the genuine non-Hermitian physics we are interested in. 

Note that the eigenvalues of a non-Hermitian Hamiltonian will generically be complex and its right and left eigenvectors will differ \cite{Brody2014,Ashida2020,Benderbook,Meden2023}. Below, we will return to this. 

As further elaborated on in Sect.~\ref{sec:s_b} it is interesting to investigate the critical properties of eigenstates of the non-Hermitian $H_{\rm eff}$. Needless to say, these critical properties will differ from those of the steady state reached when considering the full master equation (see above). 

We here do not touch upon the important question whether, in an extended many-body system such as a 1D spin chain, the no jump dynamics is observable in any finite time interval or not \cite{Daley2014}. Our reason for postponing such a discussion is pragmatic. The critical properties of non-Hermitian spin-1/2 chains were heavily studied in the past few years \cite{Lee2014,Pi2021,Liu2021,Wang2025,Liu2025,Turkeshi2023,Zhang2013a,Zhang2013b,Miao2024,Zhang2025} but a coherent picture is still lacking partly because of the use of improper methods. We here aim at a comprehensive study to clarify the somewhat fuzzy state of the field. In particular, we derive a large number of analytical results not discussed before.      

We note in passing that Eqs.~\eqref{master_nojump} and \eqref{non_herm_SG} can also be obtained within the ancilla approach \cite{Liu2019}; for a recent review see \cite{Meden2023}. In comparison to the derivation employing the Lindblad master equation including the no-jump assumption the order of steps is reversed. One starts with the quantum system with its dynamics given by the Schr\"odinger equation \eqref{non_herm_SG} with a non-Hermitian Hamiltonian. It is then embedded into a larger system by adding a single spin-1/2 degree of freedom with the dynamics of the combined system governed by a Hermitian Hamiltonian. The ancilla spin is under constant observation and only instance in which it points upwards are kept (post selection). Under this constraint the dynamics of the system with its non-Hermitian Hamiltonian and that of the system coupled to the ancilla spin with its Hermitian Hamiltonian become equivalent; for more see \cite{Meden2023}.     

\subsection{Non-Hermitian XY models}

We here consider the critical properties of two non-Hermitian XY models with magnetic field. The first is the complex-$\lambda$ (magnetic field) model \cite{Lee2014,Pi2021,Liu2021,Wang2025,Liu2025,Turkeshi2023}. This is obtained employing the Lindblad master equation in the no-jump limit when taking Eq.~\eqref{Ham} as $H_{\rm s}$, the jump operator $L_l = \sigma_l^-$, and the rate $\kappa_l= 2\I(\lambda)$ \cite{Huang2026,Konar2024,Konar2026}. Here $\sigma_l^-$ denotes the spin-1/2 lowering operator on site $l$. The effective Hamiltonian $H_{\rm eff}$ entering Eqs.~\eqref{master_nojump} and  \eqref{non_herm_SG} is then given by Eq.~\eqref{Ham} with $\lambda \in \mathbb C$ (and $\gamma \in \mathbb R)$ \cite{footnote1}.  
The model becomes non-Hermitian.

The other model is the purely imaginary $\gamma$ model \cite{Zhang2013a,Zhang2013b,Miao2024,Zhang2025}. It follows from the isotropic limit of Eq.~\eqref{Ham}, with the jump operators $ L_l=\sigma_{l+1}^{-}+\sigma_{l}^{+}$, and the rate $\kappa_l=\gamma$ \cite{Huang2026,Konar2024,Konar2026}, with $\sigma_{l}^{+}$ being the spin-1/2 raising operator on site $l$. The effective non-Hermitian Hamiltonian is then Eq.~\eqref{Ham} with $\gamma \to \mi \gamma$ (and a real $\lambda \in \mathbb R$) \cite{footnote1}. As further elaborated on below this model shows an interesting anti-unitary symmetry \cite{Zhang2013a,Zhang2013b,Miao2024,Zhang2025}. 


Several other, more complex, non-Hermitian spin-1/2 models have been studied in the literature. For example systems with so-called $\Gamma$-interaction \cite{Yan2026,Huang2026,Agarwal2025}, with Dzyaloshinskii–Moriya interaction \cite{Zhang2026}, with alternating magnetic fields \cite{Agarwal2024, Lakkaraju2021}, and with a longitudinal field \cite{Zhou2025}. It might be interesting to (re-)consider the critical properties of some of these models along the lines followed here. 

\subsection{Standard versus biorthogonal quantum mechanics}
\label{sec:s_b}

In the literature one finds two \cite{footnote2} conceptually distinct ways to define observables and expectation values in non‑Hermitian quantum systems: the biorthogonal approach \cite{Brody2014} and standard quantum mechanics. As extensively discussed in Ref.~\cite{Meden2023} biorthogonal quantum mechanics is mathematically elegant \cite{Brody2014,Benderbook} but its use for open quantum system cannot be justified on physical grounds. As was already illustrated in several examples it furthermore can lead to results which resist any meaningful physical interpretation \cite{Meden2023}. Here we provide further evidence of this for many-body spin-1/2 chains.

We thus favor the standard formalism in which an observable is a Hermitian operator $O$ and the expectation value in a give state $\left| \psi \right>$ is defined as
\begin{align}
\left< O \right> = \frac{\left< \psi \right| O \left|\psi \right>}{\left< \psi \right. \left| \psi \right>} \label{exp_val}.
\end{align}
If the state depends on time, normalization becomes crucial as the time evolution with the operator $\exp \left\{ - \mi H_{\rm eff} t \right\} $ (with time-independent $H_{\rm eff}$) is non-unitary. Due to this, expectation values of time evolved right eigenstates $\left| E \right>$ of the non-Hermitian Hamiltonian $\left| \psi(t) \right> = e^{-\mi E t } \left| E \right>$ [being a solution of Eq.~\eqref{non_herm_SG}] become stationary; any exponential increase or decrease due to a finite $\I E$ is canceled. The same holds for static correlation functions such as $C_r^\alpha$. In that sense a system prepared in one of its Hamiltonian eigenstates is stationary. It is thus meaningful to study the critical properties of such eigenstates. For reasons which become clear in the next paragraph one often refers to $\left< O \right>_{\rm RR} = \frac{\left< E \right| O \left|E \right>}{\left< E \right. \left| E \right>} $ as the right-right expectation value.    

In some of the studies \cite{Wang2025,Zhang2025} the critical properties of non-Hermitian spin-1/2 chains were investigated using the biorthogonal formalism. In this eigenstate expectation values are defined as 
\begin{align}
\langle O\rangle_{\rm LR}=\frac{\langle \tilde E|O|E\rangle}{\langle \tilde E |E  \rangle} ,
\end{align}
with the right $\left| E \right>$ eigenstate of the Hamiltonian and the corresponding left one  $|  \tilde E \rangle$ and $O$ being either an operator representing an observable or the operator part of a static correlation function (e.g., $\sigma_l^{\alpha}\sigma_{l+r}^{\alpha}$ in our case); one refers to this as the left-right expectation value.

To clarify the differences and collect further evidence that the standard formalism of quantum mechanics should be employed when considering open quantum systems within a non-Hermitian approach, we also employ the biorthogonal approach. In certain regions of the phase diagram, such as the paramagnetic phase of the complex-$\lambda$ model, the two approaches yield qualitatively similar behavior. However, significant discrepancies arise in other regimes, particularly within the regions of quasi-long-range order. A prominent example can be found in the imaginary‑$\gamma$ model. At $\lambda=\sqrt{1+\gamma^2}$, the standard formulation yields a well‑behaved correlator that captures the power-law decay of many-body correlations, whereas the biorthogonal correlator is ill-defined. Our results thus provide explicit evidence that the standard formulation admits a more robust physical interpretation, while the biorthogonal approach, although mathematically convenient, should be applied with caution when analyzing observables and correlation functions in non-Hermitian many-body systems \cite{Meden2023}.

\section{Spectral properties and expectation values}

\subsection{Eigenvalues and eigenstates}
\label{subsec:eigen}

Even with complex $\lambda$ and $\gamma$---which covers both non-Hermitian models of interest to us---the Hamiltonian \eqref{Ham} can be diagonalized using a Jordan-Wigner transformation \cite{Lieb1961}. In this the spin-1/2 raising operator is expressed as
\begin{align}
\sigma_j^+ = \exp\left[ -\mi\pi \sum_{j'=1}^{j-1} c_{j'}^\dagger c_{j'} \right] c_j \label{jord_wig}
\end{align}
with spinless fermion ladder operators $c_j^{(\dag)}$  on site $j$ \cite{Lieb1961}. Subsequent application of the Fourier transformation $c_j = \frac{1}{\sqrt{N}} \sum_{k} c_{k} e^{\mi jk}$ \cite{Lieb1961}, with discrete $k = 2 \pi n /N \in (- \pi, \pi]$, $n \in {\mathbb Z}$,  the Hamiltonian takes the form
\begin{align}
H = {\sum_{k}}^{'} \Bigg[ & \begin{pmatrix}
c_{k} \\
c_{-k}^{\dag}
\end{pmatrix}^{\dag}
\begin{pmatrix}
\lambda - \cos k & -\mi\gamma \sin k \\
\mi\gamma \sin k & -(\lambda - \cos k)
\end{pmatrix}
\begin{pmatrix}
c_k \\
c_{-k}^\dagger
\end{pmatrix} \nonumber \\ & -\cos k \Bigg] \label{ham1}
\end{align}
where the sum (with the prime) is restricted to $k \in (0, \pi]$.
In terms of non-Hermitian Bogoliubov quasi-particles with ladder operators
\begin{align}
\bar{\alpha}_k &= u_k c_k^\dagger + v_k c_{-k}, \quad \bar{\alpha}_{-k} = -v_k c_k + u_k c_{-k}^\dagger,\no\\
\alpha_k &= u_k c_k - v_k c_{-k}^\dagger, \quad \alpha_{-k} = v_k c_k^\dagger + u_k c_{-k}, 
\end{align}
Eq.~\eqref{ham1} becomes
\begin{align}
H =& -{\sum_{k}}^{'} \left(E_k+\cos k \right) + {\sum_{k}}^{'} E_k \left( \bar{\alpha}_k \alpha_k + \bar{\alpha}_{-k} \alpha_{-k} \right),\label{fh}
\end{align}
where
\begin{align}
E_k=&\pm \sqrt{(\lambda - \cos k)^2+(\gamma \sin k)^2},\label{e}
\end{align}
denotes the energy of the quasi-particles. The (normalized) Bogoliubov coefficients are 
\begin{equation}
u_k=\frac{\lambda - \cos k+E_k}{2E_ku_k}, \quad v_k=\frac{\mi\gamma \sin k}{2E_ku_k}.\label{uv}
\end{equation}
For complex $\lambda$ and/or $\gamma$  we take the standard complex root  with the branch cut along the negative real axis in Eq.~\eqref{e}. 

We emphasize that for each mode with discrete $k$ one has to select either the plus or the minus sign in Eq.~\eqref{e}. In the Hermitian model one usually takes the plus sign for all modes $k$. This way the quasi-particle vacuum state (of the Hermitian as well as the non-Hermitian model)
\begin{align}
   \left| \mathrm{Vac} \right> = {\prod_{k}}^{'}\alpha_{k}\alpha_{-k}|0 \rangle= {\prod_{k}}^{'} \left( u_k + v_k \hat{c}_k^\dagger \hat{c}_{-k}^\dagger \right) |0\rangle,\label{com_gr} 
\end{align}
with the $c$-fermion vacuum $|0\rangle$ becomes the eigenstate with lowest energy, i.e., the ground state. This is the state to consider when investigating the quantum critical properties of the Hermitian model (with $\lambda$, $\gamma \in \mathbb R$). The excited (eigen-)states are given as the occupation number states of the $\alpha$-quasi-particles. For other choices of the sign  an occupation number state with a finite number of quasi-particles will become the ground state. A proper selection of the signs in Eq.~\eqref{e} is thus a matter of convenience not of physics.  

For the complex spectrum of a generic non-Hermitian model, one does not have a clear notion of the ground state and thus no canonical way of selecting the mode signs in Eq.~\eqref{e} exists. We emphasize that this issue is ignored in many studies on non-Hermitian Hamiltonians. This observation raises the rather fundamental question which state to consider when aiming at quantum critical properties of non-Hermitian systems. As elaborated on above, expectation values and static correlation functions in right eigenstates of the non-Hermitian Hamiltonian are (at least) stationary (time independent) and such states are thus candidates. 

In fact, in the literature on the critical properties of non-Hermitian XY models two choices of eigenstates were considered. To make direct contact with the Hermitian model the quasi-particle right vacuum with the plus sign for all mode energies given in Eq.~\eqref{com_gr} was taken \cite{Liu2021,Wang2025,Miao2024}. This is the state with minimal real part of the energy; we refer to it as the minimal energy state. The corresponding left vacuum state is given by
\begin{align}
\langle \widetilde{\mathrm{Vac}}|&=\langle \tilde 0|{\prod_{k}}^{'}\bar{\alpha}_{-k}\bar{\alpha}_{k}=\langle \tilde 0| {\prod_{k}}^{'} \left( u_k - v_k \hat{c}_{-k}\hat{c}_k \right).\label{com_gl}
\end{align}

In other works the vacuum state with the signs selected such that the imaginary part of $E_k$ Eq.~\eqref{e} is always negative was taken \cite{Lee2014,Pi2021,Turkeshi2023}. This selection implies that the vacuum state has the largest imaginary part of all eigenstates. If the initial state is a superposition of eigenstates including this one, the vacuum will be the state dominating the expectation value Eq.~\eqref{exp_val} in the long-time limit (assuming that $\left< {\mathrm{Vac}} \right |O \left| {\mathrm{Vac}}  \right> \neq 0 $) which approaches a constant. For this reason it is often referred to as the steady state. Despite the fact, that according to our definition of expectation values Eq.~\eqref{exp_val}, all eigenstates of the Hamiltonian are stationary, we here also adopt this name. We will investigate the critical properties of both these eigenstates and show that they differ. 

Note that in the following both these states are given as the quasi-particle vacuum. The difference is coded in the different selection of the signs in Eq.~\eqref{e} which leads to different Bogoliubov coefficients via Eq.~\eqref{uv}. Proceeding this way has the clear advantage that all analytical expressions of Sects.~\ref{subsec:mag} and \ref{subsec:correl} apply to the minimal energy as well as to the steady state.

\subsection{The energy}
\label{subsec:ener}

From Eqs.~\eqref{fh} and \eqref{e} the energy of the vacuum state in both the right-right (RR; standard quantum mechanics) as well as the left-right (LR; biorthogonal quantum mechanics) formalism is given by 
\begin{equation}
E_{\mathrm{Vac}}=-{\sum_{k}}^{'} \left(E_k+\cos k \right).
\end{equation}
In the thermodynamic limit $N \to \infty$ the energy density $e_{\mathrm{Vac}}$ thus becomes
\begin{equation}
e_{\mathrm{Vac}}=-\frac{1}{\pi} \int_0^\pi \left(E_k+\cos k \right) \dd k.
\end{equation}
As usual in critical systems the parameter dependence of the energy density can be used to identify phase transitions. To enhance non-analytic behavior of $e_{\mathrm{Vac}}$ one can take derivatives with respect to the parameters. Recall that for non-Hermitian Hamiltonians  $e_{\mathrm{Vac}}$ will generically be complex.

\subsection{The magnetization}
\label{subsec:mag}
The magnetization per site in $z$-direction $M$ is a fundamental observable for characterizing phases and phase transitions in Hermitian spin systems. We generalize its definition by employing both RR and LR  expectation values. In the Jordan-Wigner fermions, the magnetization is expressed as $M=\sum_l\langle\sigma_l^z\rangle/N=\sum_l\langle(1-2c_l^{\dagger}c_l)\rangle/N$, where $\langle \ldots \rangle$ stands for the RR or the LR vacuum state expectation value. Employing Eqs.~\eqref{com_gr} and \eqref{com_gl} and taking the thermodynamic limit $N \to \infty$ the corresponding RR and LR expectation values are given by
\begin{align}
M_{\rm RR}&=\frac{1}{\pi} \int_0^\pi \dd k \frac{|u_k|^2 - |v_k|^2}{|u_k|^2 + |v_k|^2},\label{rr_m}\\
M_{\rm LR}&=\frac{1}{\pi} \int_0^\pi \dd k \frac{u_k^2 + v_k^2}{u_k^2 - v_k^2}.\label{lr_m}
\end{align}
From the above, it is evident that the magnetization obtained using the RR expectation value is real, whereas the LR formulation yields a complex quantity. One is thus forced to conclude that the LR magnetization is not a physical observable (see also Ref.~\cite{Meden2023}). 

\subsection{Spin correlation functions}
\label{subsec:correl}
Beyond local observables such as the magnetization, correlation functions offer a more detailed characterization and classification of phases and phase transitions. We here investigate the two-point correlation functions $C^{\alpha}_r$ Eq.~\eqref{correl_fun} along both the $\alpha=x$- and $y$-directions. 

By introducing the operators $A_l=c_l^{\dagger}+c_l$ and $B_l=c_l^{\dagger}-c_l$ \cite{Lieb1961}, the correlation functions can be expressed as
\begin{eqnarray}
C^x_r&=& \langle B_l A_{l+1} B_{l+1} \cdots A_{l+r-1} B_{l+r-1} A_{l+r} \rangle,\\
C^y_r&=& (-1)^r \langle A_l B_{l+1} A_{l+1} \cdots B_{l+r-1} A_{l+r-1} B_{l+1} \rangle .
\end{eqnarray}
Using Wick's theorem, $C^{x}_r$ can be written as the Pfaffian of a $2r\times2r$ block Toeplitz matrix
\begin{equation}
C^x_r=\mathrm{Pf}
\begin{bmatrix}
M_0   & M_1   & \cdots & M_{r-1}   \\
M_{-1}   & M_0   & \cdots & M_{r-2} \\
\vdots & \vdots & \ddots & \vdots \\
M_{1-r} & M_{2-r} & \cdots & M_0
\end{bmatrix},\label{pf}
\end{equation} 
where $M_{-r}=-M_r^\mathrm{T}$ and $M_r$ is a $2 \times 2$ matrix
\begin{equation}
M_r=\begin{bmatrix}
\langle B_0 B_{r} \rangle +\delta_{0r} & \langle B_0 A_{r+1} \rangle   \\
\langle A_0 B_{r-1} \rangle &\langle A_0 A_{r} \rangle-\delta_{0r}
\end{bmatrix},\label{M}
\end{equation}
denoting the two-point correlations. Utilizing that the Pfaffian of a skew-symmetric matrix is equivalent to the square root of its determinant, the correlation functions can be simplified. 

Specifically, for the RR expectation value, the elements of the matrix $M_r$ are  
\begin{align}
\langle A_0 A_{r} \rangle_{\rm RR} &= \delta_{0r} + \frac{\mi}{\pi} \int_0^\pi \dd k \frac{u_k v_k^* + u_k^* v_k}{|u_k|^2 + |v_k|^2}  \sin(kr),\\
\langle B_0 B_{r} \rangle_{\rm RR} &= -\delta_{0r} + \frac{\mi}{\pi} \int_0^\pi \dd k \frac{u_k v_k^* + u_k^* v_k}{|u_k|^2 + |v_k|^2}  \sin(kr),\\
\langle B_0 A_{r} \rangle_{\rm RR} &= -\langle A_{r} B_0 \rangle_{\rm RR}\no \\ 
&= \frac{1}{\pi} \int_0^\pi \dd k \frac{ |v_k|^2-|u_k|^2}{|u_k|^2 + |v_k|^2} \cos(kr) \no\\ &+ \frac{1}{\pi} \int_0^\pi \dd k \frac{u_k v_k^* - u_k^* v_k}{|u_k|^2 + |v_k|^2} \mi\sin(kr).
\end{align}

For the LR expectation value, the elements of the matrix $M_r$ are
\begin{eqnarray}
\langle A_0 A_{r} \rangle_{\rm LR} &=& \delta_{0r},\\
\langle B_0 B_{r} \rangle_{\rm LR} &=& -\delta_{0r},\\
\langle B_0 A_{r} \rangle_{\rm LR} &=& -\langle A_{r} B_0 \rangle_{\rm LR} \no\\
&=& -\frac{1}{\pi} \int_0^\pi \dd k \frac{ v_k^2+u_k^2}{u_k^2 - v_k^2} \cos(kr) \no\\&&- \frac{1}{\pi} \int_0^\pi \dd k \frac{2u_k v_k}{u_k^2 - v_k^2} \mi\sin(kr).\label{elr}
\end{eqnarray}
It can be seen that the diagonal elements of the matrix $M_r$ vanish in the LR formulation. Therefore, the Pfaffian in Eq.~\eqref{pf} can be reduced to the determinant of a $r\times r$ scalar Toeplitz matrix
\begin{equation}
\left(C^x_r\right)_\mathrm{LR}=
\begin{vmatrix}
\tilde{M}_0   & \tilde{M}_1   & \cdots & \tilde{M}_{r-1}   \\
\tilde{M}_{-1}   & \tilde{M}_0   & \cdots & \tilde{M}_{r-2} \\
\vdots & \vdots & \ddots & \vdots \\
\tilde{M}_{1-r} & \tilde{M}_{2-r} & \cdots & \tilde{M}_0
\end{vmatrix},\label{det}
\end{equation}
where $\tilde{M}_{r}=\langle B_0 A_{r+1} \rangle$ is the element in the first row and second column of the matrix $M_{r}$. 

\begin{figure*}[t] 
	\begin{center} 
		\includegraphics[width=0.93\linewidth]{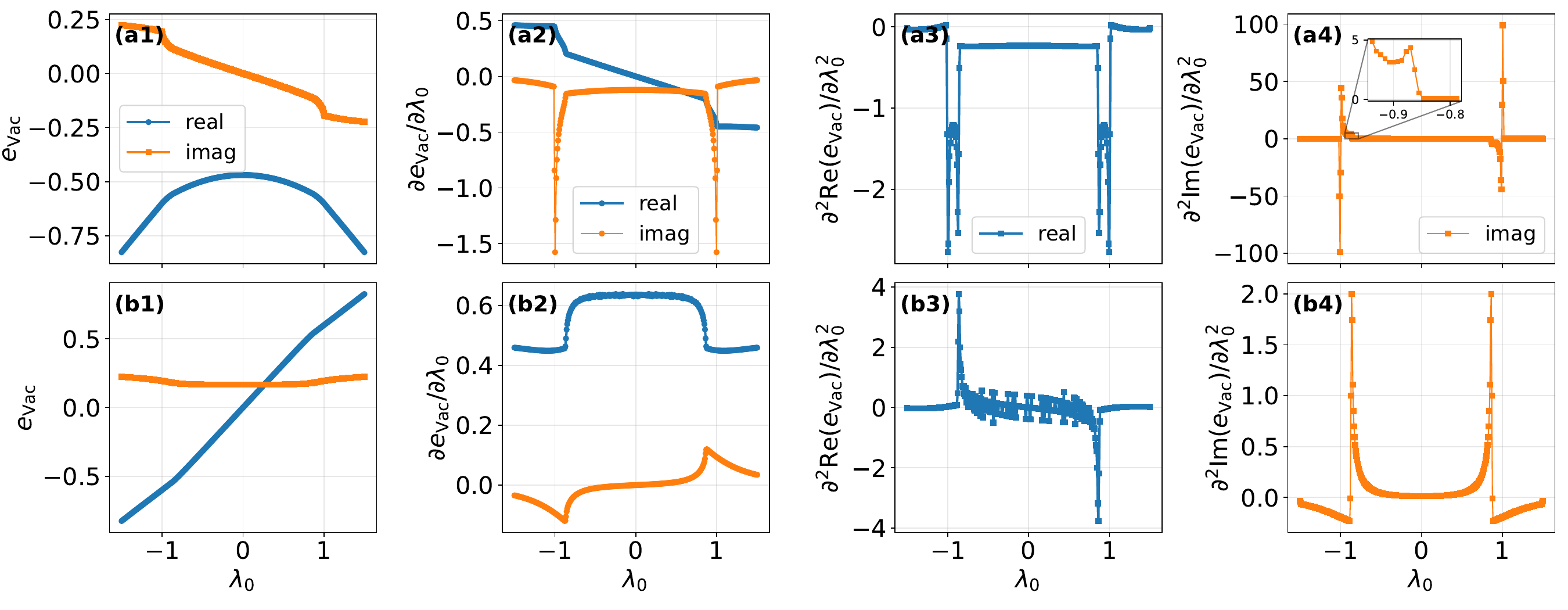} 
	\end{center}
	\vspace{-14pt}  
	\caption{The vacuum state energy density as a function of $\R(\lambda) = \lambda_0$ for the minimal energy state (a) and the steady state (b) at $\I(\lambda)=1/2$ and $\gamma=1$. From left to right, the energy density, its first derivative with respect to $\lambda_0$, and the second derivative of its real and imaginary part are shown.} 
	\label{com_gs_energy}
\end{figure*}

The correlation function in $y$-direction, $C^y_r$, can be directly obtained from Eqs.~\eqref{pf} and \eqref{det} by multiplying by an overall factor $(-1)^r$ and replacing the correlators  $\langle O_0 O'_r \rangle$ with $-\langle O_0 O'_{-r} \rangle$ for $O, O' \in \{A, B\}$. 

From these expressions, it is evident that the RR correlation function remains real, whereas the LR correlation function is intrinsically complex. Equations \eqref{pf} and \eqref{det} are universal and can be applied to other non-Hermitian XY models once the Bogoliubov coefficients are known. 

\section{Complex-$\lambda$ model}

In this section, we consider the Hamiltonian \eqref{Ham} with a complex magnetic field $\lambda$ but $\gamma \in \mathbb R$ \cite{Lee2014,Pi2021,Liu2021,Wang2025,Liu2025,Turkeshi2023}. We provide a systematic analysis of the energy, the magnetization, and the correlation functions for the minimal energy and the steady state, introduced in Sect.~\ref{subsec:eigen}. For the magnetization and the correlation functions we employ both, the standard as well as the biorthogonal formalism.

\subsection{The energy and energy gaps}
\label{sec:eandg}

In Fig.~\ref{com_gs_energy}, we plot the vacuum state energy density $e_{\mathrm{Vac}}$, and its first as well as second derivatives as a function of $\R(\lambda) = \lambda_0$ for the minimal energy (a) as well as the steady state (b) at $\I(\lambda)=1/2$ and $\gamma=1$ . The real and imaginary parts of the energy density are continuous as shown in (a1) and (b1). Hints on possible non-analytic behavior, and thus phase transitions, are barely visible. In contrast, the first derivatives in (a2) and (b2) exhibit pronounced non-analyticities indicative of critical points. These are further enhanced in the second derivatives (a3), (a4), (b3), and (b4). 

While the energy of the minimal energy state in the upper row of Fig.~\ref{com_gs_energy} indicates four transitions when varying $\R(\lambda)$ at fixed $\I(\lambda)$ and $\gamma$, the energy of the steady state, shown in the lower row, signals only two transitions. We can thus already now conclude that the critical properties depend on the state considered. What is measured in an (gedanken-)experiment would thus depend on the state in which the system is prepared. We here exclusively consider the two cases of the minimal energy eigenstate of the non-Hermitian Hamiltonian with the smallest real part of the energy and the eigenstate with the largest imaginary part of the energy which dominates expectation values in the large time limit. Considering other eigenstates might be meaningful.       

Another interesting observation from Fig.~\ref{com_gs_energy} (a1) is that for a purely imaginary magnetic field (i.e., $\lambda_0=0$), the energy of the minimal energy state is real. In addition, if eigenstates are formed by creating two quasi-particles with opposite imaginary parts of the energy, the total energy of these states also remains real. This demonstrates that non-Hermitian systems can host eigenstates with real energies even without explicit (anti-unitary) global symmetries, such as parity-time \cite{Benderbook,Meden2023} or $\mathcal{RK}$ (see below) symmetry.

The opening and closing of energy gaps are routinely used to detect critical points in Hermitian systems \cite{Sachdev2011}. For the minimal energy and the steady state in our non-Hermitian complex-$\lambda$ model different gaps matter according to the definition of the respective states. For the minimal energy state we define the energy gap as the quasi-particle excitation energy evaluated at the momentum where the real part of $E_k$ Eq.~\eqref{e} reaches its minimum, i.e., $\Delta_{\rm min}= E_k \big|_{\min_k\R(E_k)}$ \cite{Liu2021}. This is in full accordance with the gap definition in the Hermitian XY model with (real) magnetic field. In contrast, for the steady state, the relevant energy gap is defined as the quasi-particle energy at the momentum where the imaginary part of $E_k$ reaches its maximum, i.e., $\Delta_{\rm ss}=E_k \big|_{\max_k\I(E_k)}$. 

\begin{figure}[tbp] 
	\begin{center} 
		\includegraphics[width=0.7\linewidth]{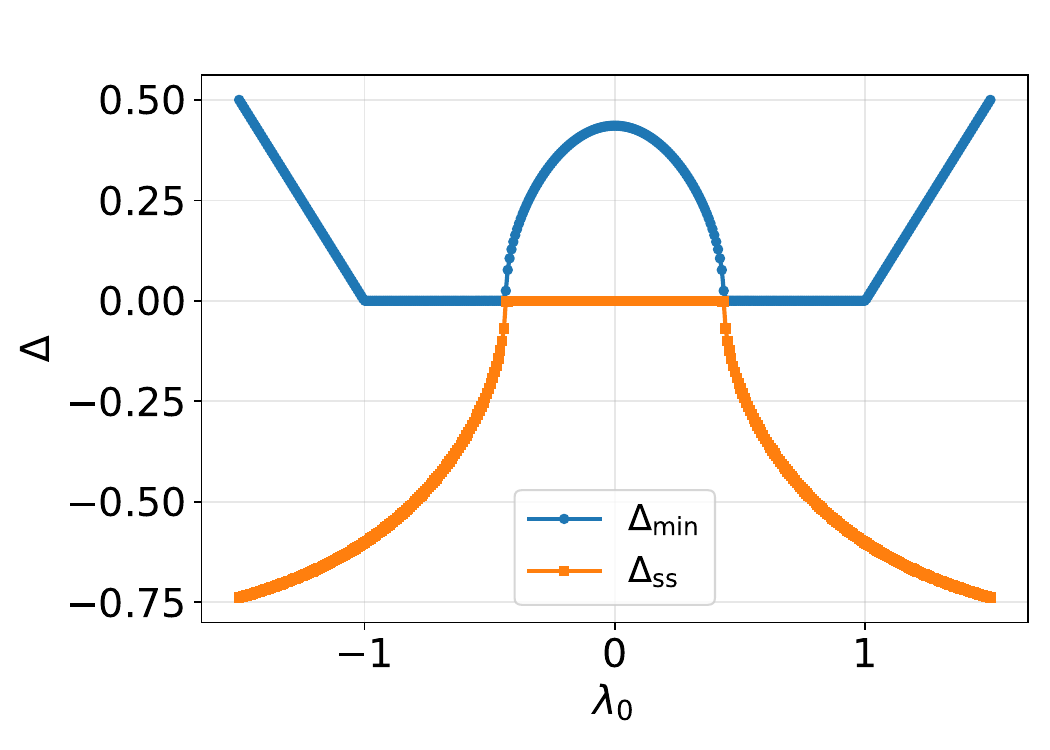} 
	\end{center}
	\vspace{-14pt}  
	\caption{The relevant energy gaps $\Delta$ for the minimal energy state (blue) and the steady state (orange) as a function of $\R(\lambda) = \lambda_0$ at $\I(\lambda)=9/10$ and $\gamma=1$.}         
	\label{en_min}
\end{figure}

\begin{figure}[t] 
	\begin{center} 
		\includegraphics[width=0.98\linewidth]{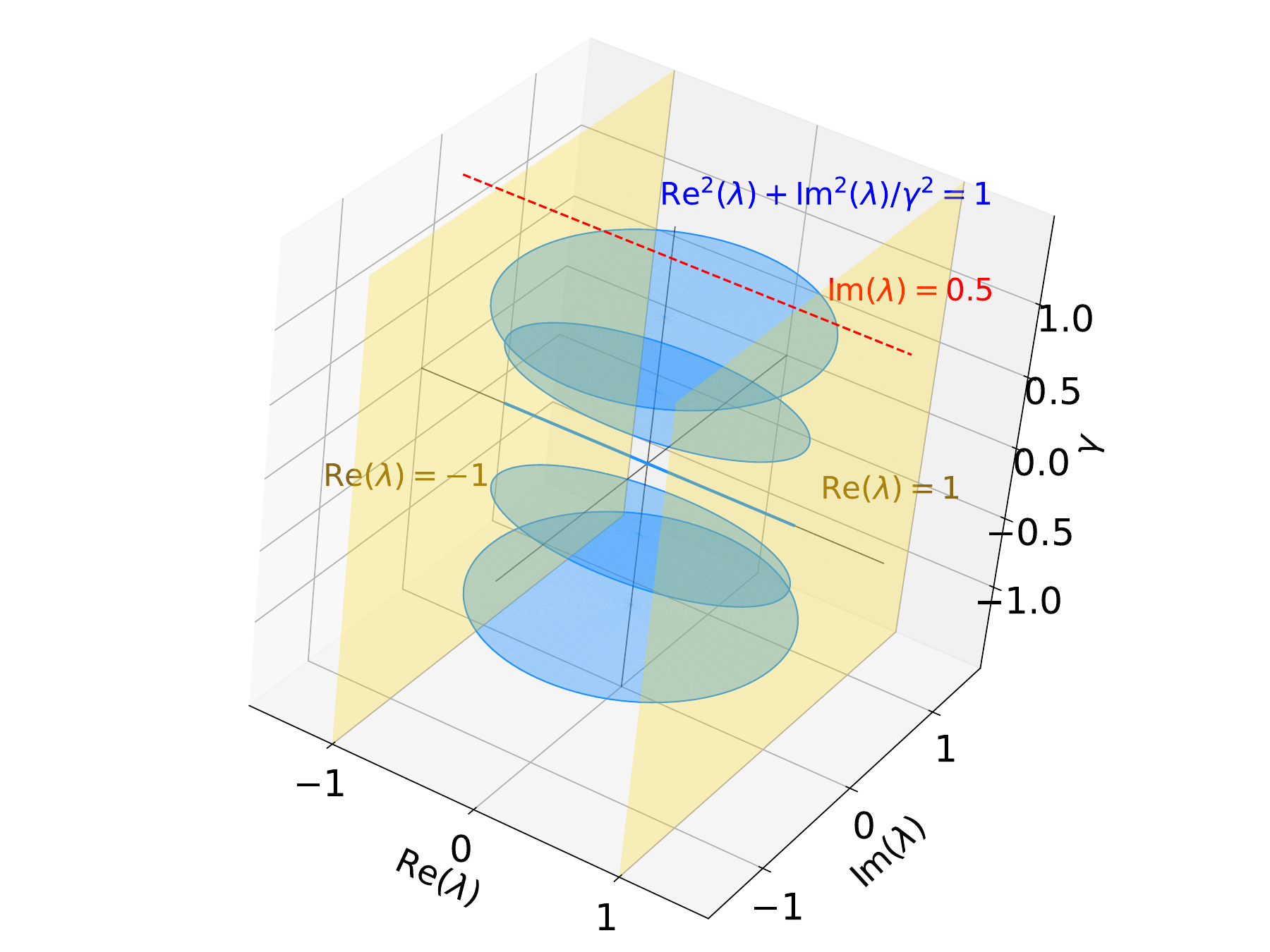}
	\end{center}
	\vspace{-14pt}  
	\caption{The three-dimensional phase diagram of the complex-$\lambda$ XY model as a function of $\R(\lambda),\I (\lambda)$ and $\gamma$ for the minimal energy state. The blue surfaces indicate the FM phase for selected $\gamma$. The yellow sheets indicate the transition to the paramagnetic phase with $|\R(\lambda)|>1$. In the inner region between the yellow sheets and outside the blue areas the state shows LL critical properties.
	The red line depicts the path used in the upper row of  Fig.~\ref{com_gs_energy}.}    
	\label{pd_3D}
\end{figure}

The dependence of these two gaps on $\R(\lambda) = \lambda_0$ at $\I(\lambda)=9/10$ and $\gamma=1$ is shown in Fig.~\ref{en_min}. According to Eq.~\eqref{e} both gaps vanish at the critical point $\lambda_{\rm c}$, determined by $\R^2 (\lambda_{\rm c})+\I^2(\lambda_{\rm c})/\gamma^2=1$. However, the gapless regions differ between the two choices of state. For the minimal energy state at fixed $\gamma$, the quasi-particle spectrum is gapless in the regime $\lambda_{\rm c}<|\lambda|<1$, whereas for the steady state it is gapless for $|\lambda|<\lambda_{\rm c}$. The gapless regimes correspond to critical phases, in which the correlation functions exhibit power-law decay; see below. A further difference lies in the structure of the phase diagram. For the minimal energy state three distinct phases can be found, while the steady state yields only two phases. Specifically, the phase boundaries are given by
\begin{eqnarray}
&&\text{  FM: } \R^2 (\lambda)+\frac{\I^2(\lambda)}{\gamma^2}<1,\no\\
&&\text{   LL: } \R^2 (\lambda)+\frac{\I^2(\lambda)}{\gamma^2}>1 \text{ and } |\R (\lambda)|<1,\no\\
&&\text{   PM: } |\R (\lambda)|>1
\label{phases_min_energ}
\end{eqnarray}
for the minimal energy state and
\begin{eqnarray}
&&\text{   LL: } \R^2 (\lambda)+\frac{\I^2(\lambda)}{\gamma^2}<1,\no\\
&&\text{   PM: }\R^2 (\lambda)+\frac{\I^2(\lambda)}{\gamma^2}>1
\end{eqnarray}
for the steady state. We already indicated the phases as FM, LL, and PM as the properties of the correlation functions discussed in Sect.~\ref{sec:correl} will show a strong (but not a full) analogy to the corresponding phases of the Hermitian XY model with magnetic field.

\begin{figure*}[tbp] 
	\begin{center} 
		\includegraphics[width=0.96\linewidth]{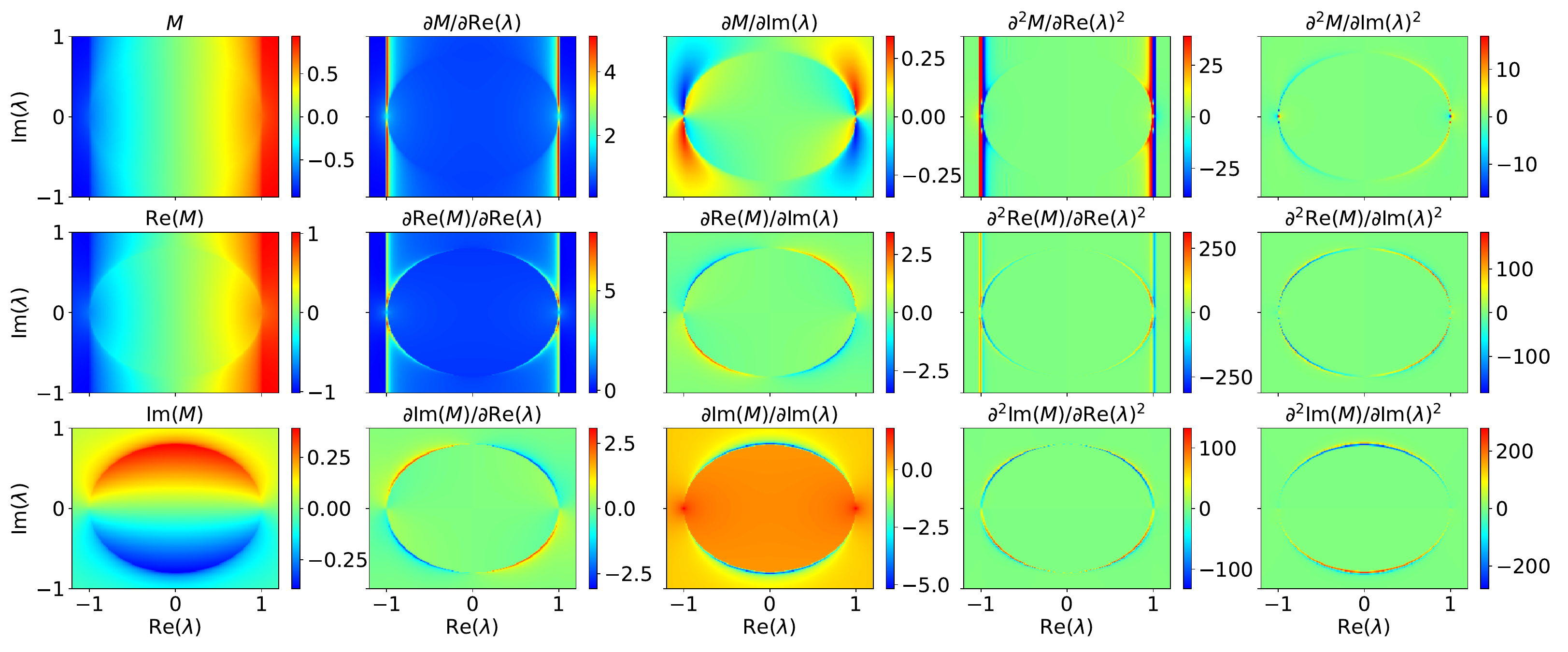} 
	\end{center}
	\vspace{-14pt}  
	\caption{The magnetization of the minimal energy state as a function of the real and imaginary part of $\lambda$ for $\gamma=0.8$ and its first and second derivative with respect to the two variables. From top to bottom: the RR magnetization, the real  and imaginary parts of the LR magnetization.}         
	\label{com_mag1}
\end{figure*}

\begin{figure*}[tbp] 
	\begin{center} 
		\includegraphics[width=0.96\linewidth]{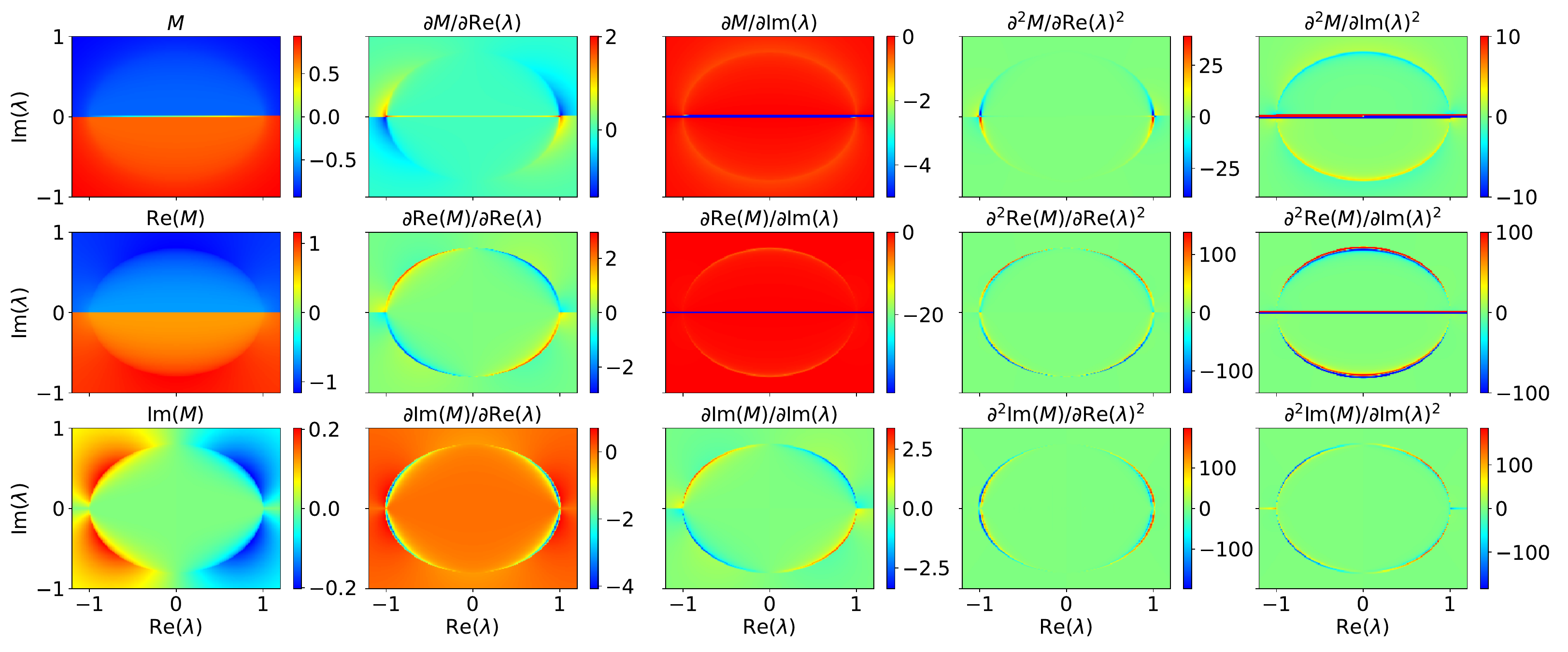} 
	\end{center}
	\vspace{-14pt}  
	\caption{The same as in Fig.~\ref{com_mag1} but for the steady state.}         
	\label{com_mag2}
\end{figure*}

The (more interesting) phase diagram for the minimal energy state as a function of the three parameters $\R(\lambda)$, $\I(\lambda)$ and $\gamma$ is shown in Fig.~\ref{pd_3D}. The blue surfaces indicate the FM phase for selected fixed $\gamma$. At $|\gamma|=1$ (upper and lower blue surfaces) this is given by a circle of radius 1. For $-1 < \gamma < 1$ this circle is squeezed to an ellipse in the $\I(\lambda)$-direction. The yellow surfaces, defined by $|\R(\lambda)|=1$, indicate the transition to the paramagnetic phase for which $|\R(\lambda)|>1$. In the inner region between the yellow sheets and outside the blue areas the state shows LL critical properties.  

\subsection{Magnetization}

In this section, we analyze the behavior of the magnetization both within and across different phases. 

In Fig.~\ref{com_mag1}, we plot the magnetization and its first and second derivatives with respect to $\R(\lambda)$ and $\I(\lambda)$ at fixed $\gamma=0.8$ for the minimal energy state. Note the different scales of the color coded magnetization. The first row shows the (real) RR expectation value Eq.~\eqref{rr_m}. The second row shows the same for the real part of the LR magnetization and the third the corresponding imaginary part. In all rows clear hints on the FM-LL phase transition can be seen (ellipse), while the imaginary component of the LR magnetization fails to provide signatures of the LL–PM transition. We reemphasize that the (real) magnetization obtained as the RR expectation value is physically meaningful. The complex nature of the LR magnetization in contrast resists any physically sensible interpretation.  

Figure \ref{com_mag2} shows the magnetization for the steady state. In addition to the LL-PM phase boundary (ellipse), a distinct line at $\I(\lambda)=0$ is observed. This is the limit in which the model becomes the Hermitian XY model with (real) magnetic field. In this the defining criterion of the steady state as the eigenstate with the largest imaginary part of the energy loses its meaning as the energy becomes real.

\subsection{Correlation functions}
\label{sec:correl}

Based on the Eqs.~\eqref{pf}-\eqref{det}, we now investigate the large-$r$ asymptotic behavior of the  correlation functions $C_r^{x/y}$ in the entire parameter space. In a first step we present numerical results and then discuss analytical insights. To be concise, the following discussion mainly focuses on the minimal energy state. A brief comparison with results for the steady state is provided at the end of this section.

\begin{figure}[tbp] 
	\begin{center} 
		\includegraphics[width=1\linewidth]{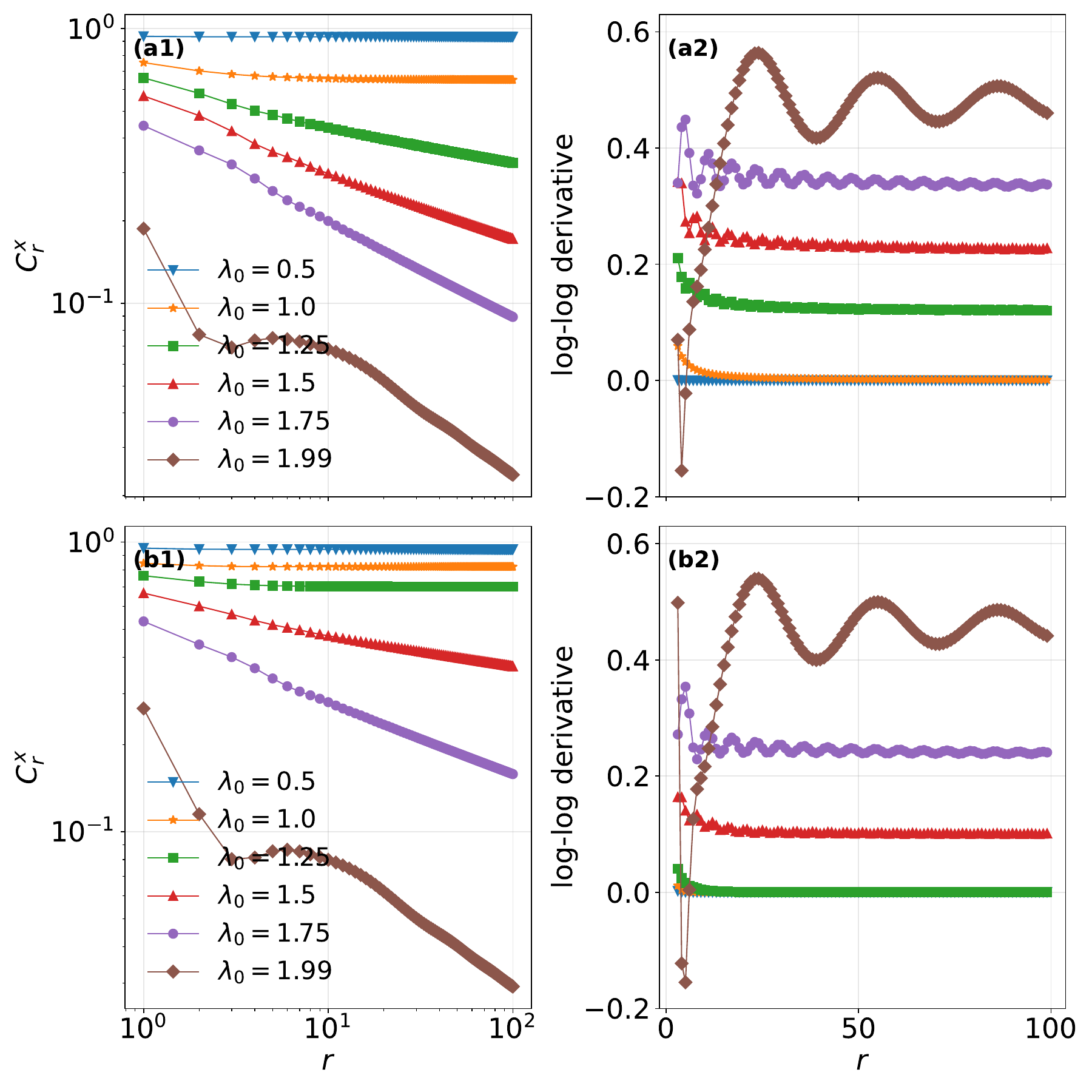} 
	\end{center}
	\caption{The $r$-dependence of the RR $x$-correlation function $C_r^x$ for different $\lambda_0$, with $\lambda=\lambda_0\exp(\mi\pi/3)$, at $\gamma=1$ (upper row) and $\gamma=1.5$ (lower row). Panels (a1) and (b1) are the bare data presented on a log–log scale. Panels (a2) and (b2) show the log-log derivative $-\dd\ln|C_r^x| / \dd\ln r$.}
	\label{com_x_rr_lh_ga}
\end{figure}

\begin{figure}[tbp] 
	\begin{center} 
		\includegraphics[width=1\linewidth]{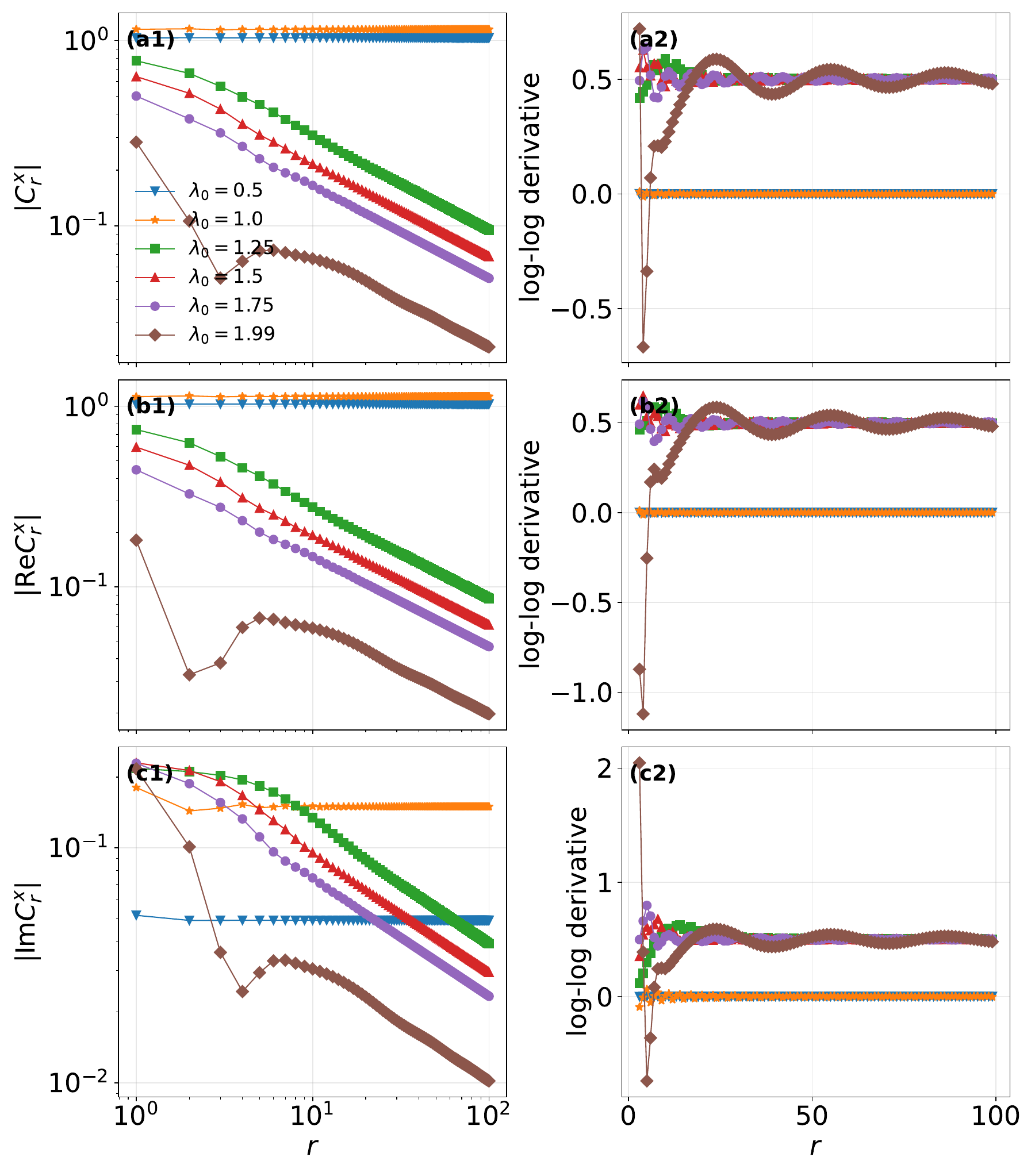} 
	\end{center}
	\vspace{-14pt}  
	\caption{The $r$-dependence of the LR $x$-correlation function $C_r^x$ for different $\lambda_0$, with $\lambda=\lambda_0\exp(\mi\pi/3)$, at $\gamma=1$. The three rows show the absolute value, the real, and the imaginary part, respectively. Panels (a1), (b1), and (c1) are the bare data presented on a log–log scale. Panels (a2), (b2), and (c2) show the log-log derivative $-\dd\ln|C_r^x| / \dd\ln r$.}         
	\label{com_x_lr_lh_ga}
\end{figure}

\subsubsection{$C_r^x$ in the FM phase and at the FM to LL transition}

The left panels of Figs.~\ref{com_x_rr_lh_ga} and \ref{com_x_lr_lh_ga} show the $r$-dependence of the RR and LR $x$-correlation function $C_r^x$ on a log-log scale, respectively. For the LR expectation value the absolute value, the real as well as the imaginary parts are shown. The magnetic field is taken as $\lambda=\lambda_0\exp(\mi\pi/3)$, where the modulus $\lambda_0 \in \mathbb R$ varies while the phase $\phi=\pi/3$ is fixed. According to Eq.~\eqref{phases_min_energ} this choice leads to a FM to LL phase transition at $\lambda_0^{\rm c}=\sqrt{4\gamma^2/(3+\gamma^2)}$. The anisotropy is set to $\gamma=1$ (upper panel) and $\gamma=1.5$ (lower panel) in Fig.~\ref{com_x_rr_lh_ga} and $\gamma=1$ in Fig.~\ref{com_x_lr_lh_ga}. Evidently, for $\gamma=1$ the blue and orange curves and for $\gamma=1.5$ the blue, orange, and green curves approach  constants for large $r$ indicative of FM order known from the Hermitian model. This explains the nomenclature we introduced in Eq.~\eqref{phases_min_energ}. In this parameter regime the gap as defined in Sect.~\ref{sec:eandg} is finite, a characteristic also known from the FM phase of the Hermitian model. Note that for $\gamma=1$ the FM to LL transition is located at $\lambda_0=1$ (orange curves in Figs.~\ref{com_x_rr_lh_ga} (a1) and \ref{com_x_lr_lh_ga} (a1)). We can thus conclude that long-range order is also found at the phase transition.   

The right panels of Figs.~\ref{com_x_rr_lh_ga} and \ref{com_x_lr_lh_ga} show the log-log derivative $-\dd\ln|C_r^x| / \dd\ln r$ of the left panel data evaluated as discrete centered differences on the lattice. In the FM phase and at the phase transition this approaches zero at large $r$ consistent with the approach to a constant of the bare data. In Sect.~\ref{sec:whylog} it becomes evident why we consider the log-log derivative in addition to the bare data. 

The RR and the LR expectation values show qualitatively the same behavior. However, the asymptotic values of the correlation function differ. 

\subsubsection{$C_r^x$ in the LL phase}
\label{sec:whylog}

For $\sqrt{4\gamma^2/(3+\gamma^2)}<\lambda_0<2$, with $\lambda_0^{\rm c}=2$ corresponding to $\R(\lambda)=1$ for $\lambda=\lambda_0\exp(\mi\pi/3)$, the log-log scale data in the left panels of Figs.~\ref{com_x_rr_lh_ga} and \ref{com_x_lr_lh_ga} appear to follow a straight line at large $r$. This is indicative of power-law scaling as known from the (gapless) LL phase of the Hermitian XY model with magnetic field. This explains the nomenclature introduced in Eq.~\eqref{phases_min_energ}. Evidently, the scale beyond which this behavior is found increases with increasing $\lambda_0$. To further investigate this, the right panels show the log-log derivative. For an asymptotic power-law scaling the log-log derivative approaches a constant, with the latter being the exponent. This is a very sensitive probe of asymptotic power-law scaling, going way beyond any fitting procedure \cite{Liu2021,Miao2024}. Indeed the curves in the LL phase all approach a constant. The closer one comes to the transition at $\lambda_0^{\rm c}=2$ the larger becomes the frequency as well as the amplitude of oscillations in the log-log derivative requiring larger and larger $r$ to find clear indications of power-law behavior. For a more detailed analysis of this, see Appendix~\ref{oscis}.

In the LL phase we find a qualitative difference between the RR and the LR correlation functions. In the former the asymptotic value of the log-log derivative and thus the exponent in the LL phase depends on $\lambda_0$ and $\gamma$ and varies between $0$ (close to the FM-LL phase boundary) and $1/2$ (when approaching the transition at $\lambda_0^{\rm c} =2$) while in the latter the exponent is always $1/2$, as known from the LL phase of the Hermitian model. We will further comment on this difference after having discussed $C_r^y$ as well.


The power-law decay of the RR and LR correlation functions in the LL phase can be understood analytically within the framework of the Fisher–Hartwig theorem, corresponding to the scalar \cite{Ehrhardt2001a,Ehrhardt2001b,Deift2011,Jin2004} and block cases \cite{Ares2015,Ares2018,Basor2025}, respectively. For a scalar Toeplitz matrix, the asymptotic behavior is by now essentially well established, with the exponent determined by the associated jump and root singularities. In contrast, for the block Toeplitz case, analytical results remain limited and are primarily restricted to specific scenarios, such as those involving jump singularities only with restricted condition. The general universal framework is more challenging and has not been established hitherto \cite{Basor2025}. In Appendix~\ref{ll_ex}, we provide an analytical derivation of the power-law exponents for both the RR and LR correlation functions. In particular, the theoretical analysis shows that the LR exponent is 1/2, which agrees with the numerical results, while the RR exponent is 
$
2(\theta_0/ \pi)^2
$
with 
\begin{eqnarray}
&&\cos \theta_0=\frac{e^{-d}-e^{d}}{e^{-d}+e^{d}}, \quad \sin \theta_0=\frac{2}{e^{-d}+e^{d}},\no\\
&&e^{d}=\left(\frac{1-m}{1+m}\right)^{1/2}, \quad m=\frac{\gamma \sqrt{1-\R^2(\lambda)}}{\I(\lambda)}.\label{theta0}
\end{eqnarray}

\begin{figure}[tbp] 
	\begin{center} 
		\includegraphics[width=0.9\linewidth]{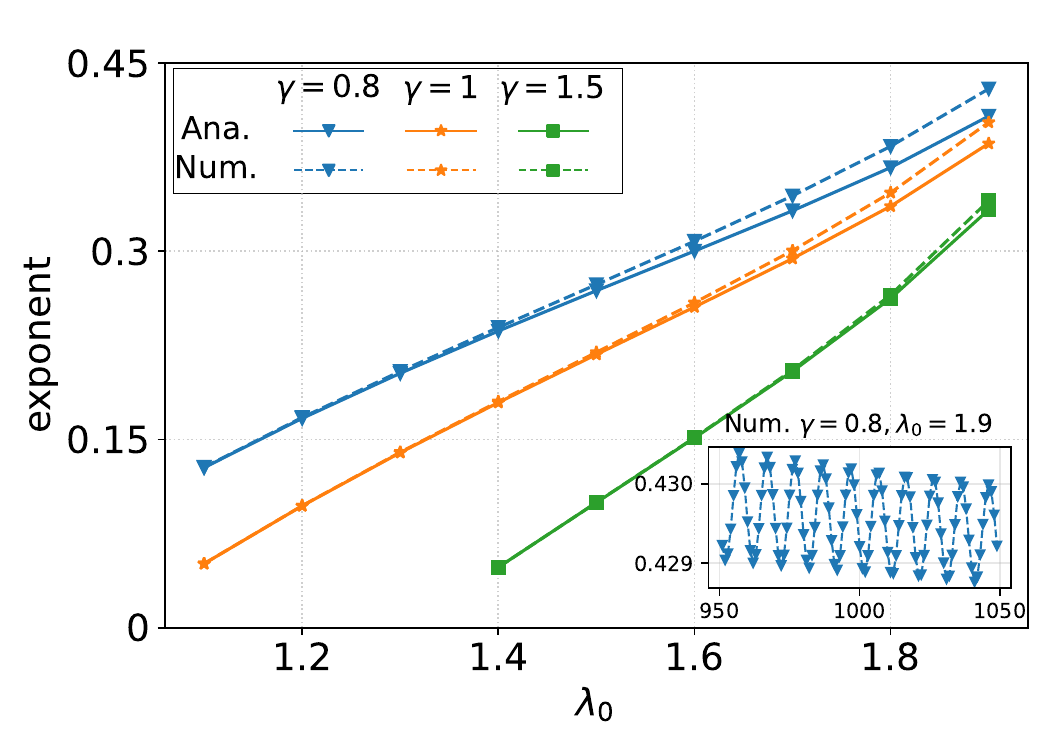} 
	\end{center}
	\vspace{-14pt}  
	\caption{Analytical and numerical results for the RR exponent in the LL phase as a function of $\lambda_0$ for $\lambda=\lambda_0\exp(\mi\pi/3)$ and different $\gamma$. The inset shows the behavior of the numerical log-log derivative of $C_r^x$ around a distance of $r \sim 1000$ for $\gamma=0.8$ and $\lambda_0=1.9$.}         
	\label{rr_ex_ll}
\end{figure}

A comparison between the analytical and numerical results for the RR exponent as a function of $\lambda_0$ and for different $\gamma$ is presented in Fig.~\ref{rr_ex_ll}. The numerical exponent is extracted from the log-log derivative of the correlation function at a distance of $r \sim 1000$ after averaging over the remaining oscillations. For smaller $\gamma$ and near the PM phase transition $\lambda_0 \nearrow 2$, deviations between the analytical and numerical exponents can be observed. These can be explained considering the inset of Fig.~\ref{rr_ex_ll}, showing that for such $\gamma$ and $\lambda_0$ the log-log derivative of the correlation function has not converged yet (even after averaging over the oscillations) and much larger $r$ must be considered. Taking this into account, the analytical and numerical results are in convincing agreement.

\begin{figure}[tbp] 
	\begin{center} 
		\includegraphics[width=1\linewidth]{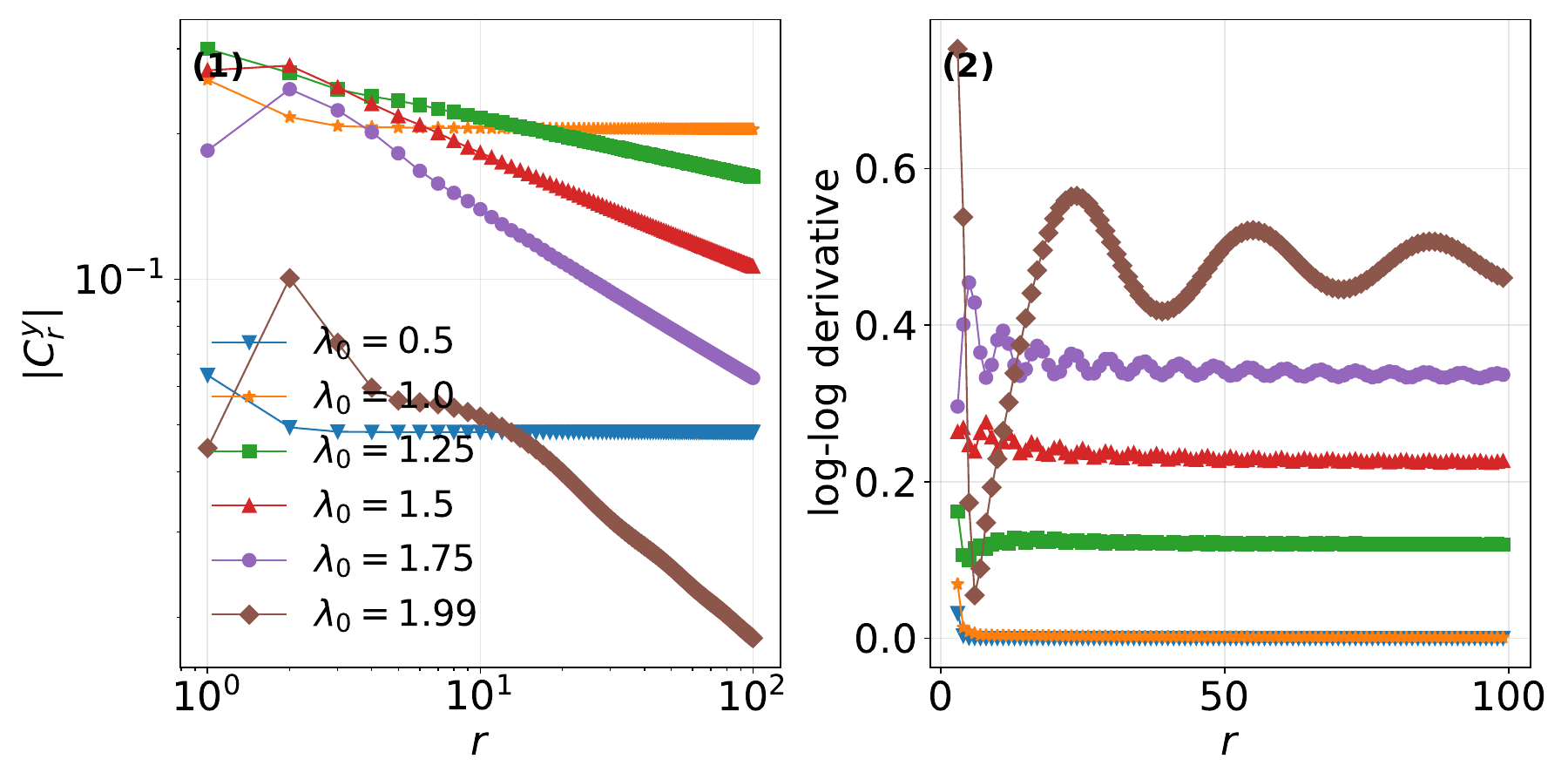} 
	\end{center}
	\caption{The $r$-dependence of the RR $y$-correlation function $C_r^y$ for different $\lambda_0$, with $\lambda=\lambda_0\exp(\mi\pi/3)$, at $\gamma=1$. Panel (1) shows the bare data presented on a log–log scale while panel (2) shows the log-log derivative $-\dd\ln|C_r^x| / \dd\ln r$.}
	\label{com_y_rr_lh_ga}
\end{figure}

\subsubsection {$C_r^y$ in the FM and LL phases}

We now turn to the correlation functions along the $y$ direction. Figure \ref{com_y_rr_lh_ga} displays the RR $y$-correlation function $C_r^y$ in the FM and LL phases. It shows the same asymptotic large $r$ behavior as $C_r^x$ of Fig.~\ref{com_x_rr_lh_ga}. In particular, both correlation functions saturate to a finite constant in the FM phase. Note that this is in difference to the FM$_x$ phase of the Hermitian model in which $C_r^y$ goes to zero exponentially; see Table \ref{tab_her}. The $x$ and $y$ correlation functions approach a finite constant also at the FM–LL transition point, whereas in the LL phase they exhibit power-law decay with identical $\lambda_0$ and $\gamma$-dependent exponents (and identical oscillation periods; see Appendix \ref{oscis}).

\begin{figure}[tbp] 
	\begin{center} 
		\includegraphics[width=1\linewidth]{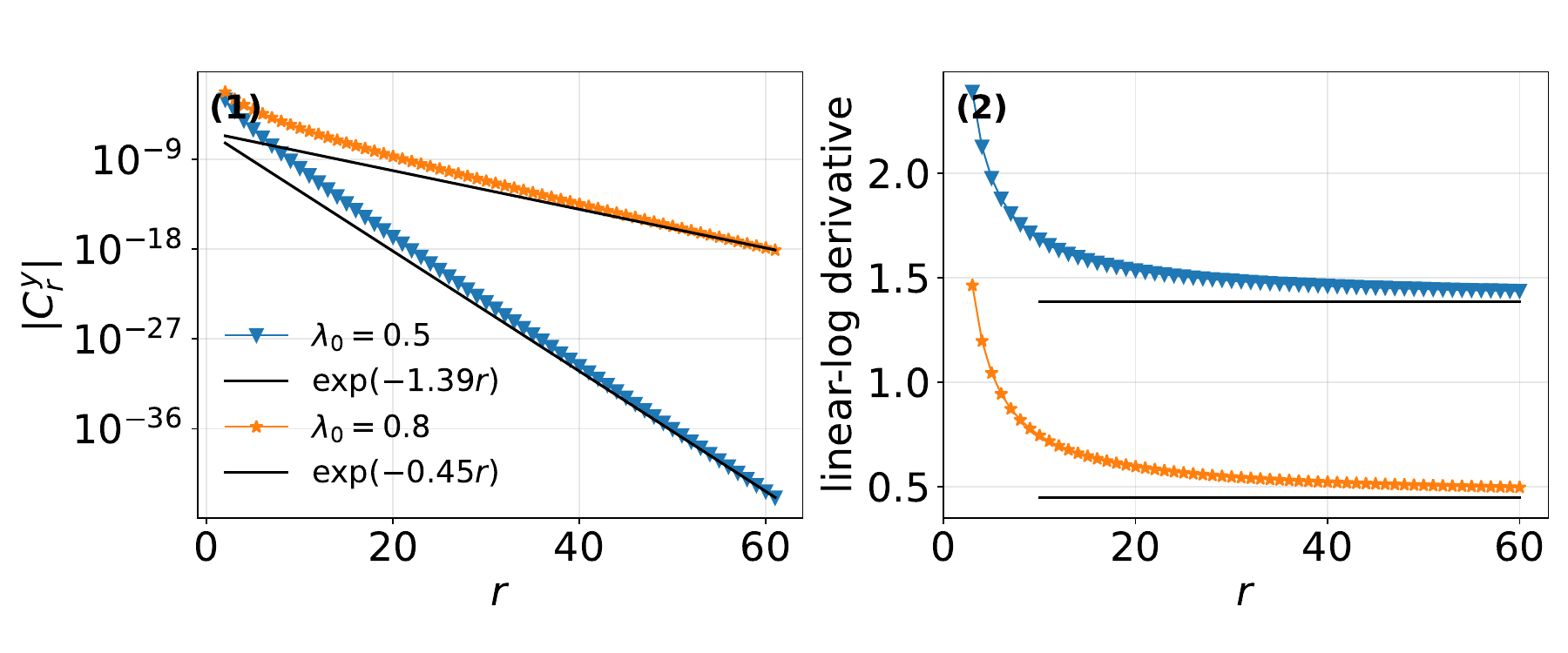} 
	\end{center}
	\vspace{-14pt}  
	\caption{The $r$-dependence of the modulus of the LR $y$-correlation function $\left|C_r^y\right|$ for different $\lambda_0$, with $\lambda=\lambda_0\exp(\mi\pi/3)$, at $\gamma=1$ in the FM phase. Panel (1) shows  the bare data presented on a linear-log scale. Panel (2) shows the linear-log derivative $-\dd\ln|C_r^x| / \dd r$. The black lines show analytical results for the inverse correlation length. Note that very small values for the correlation function can be obtained in the numerical evaluation by using appropriate data formats; see the $y$-axis scale of panel (1).}     
	\label{com_y_lr_fm_ga}
\end{figure}

Taking the LR expectation values the behavior of the correlation functions $\left|C_r^x\right|$ and $\left|C_r^y\right|$ in the FM phase clearly differ; from now on, we focus on the modulus of the complex LR correlation functions. As discussed above $\left|C_r^x\right|$  approaches a constant; see the blue and orange curves in Fig.~\ref{com_x_lr_lh_ga} (a1). In contrast, $\left|C_r^y\right|$ decays exponentially to zero as shown in Fig.~\ref{com_y_lr_fm_ga}. Note the linear-log scale of the left panel and the corresponding linear-log derivative $-\dd \ln|C_r^x| / \dd r$ (evaluated as centered differences) shown on the right. If the asymptotic decay is exponential this derivative approaches the constant inverse correlation length $\xi^{-1}$ at large $r$.  As further elaborated on in Sect.~\ref{sec:PM_phase} the correlation length can be computed analytically. The corresponding results are shown as black lines in Fig.~\ref{com_y_lr_fm_ga}. The exponential decay of the $y$-correlation function in a phase in which the $x$-correlation function approaches a constant is reminiscent of the behavior of the FM phase of the Hermitian model. In this a $\gamma>0$ favors the formation of an $x$-correlated ferromagnetic phase ($\text{FM}_x$), thereby exponentially suppressing correlations along the $y$ direction.


\begin{figure}[tbp] 
	\begin{center} 
		\includegraphics[width=1\linewidth]{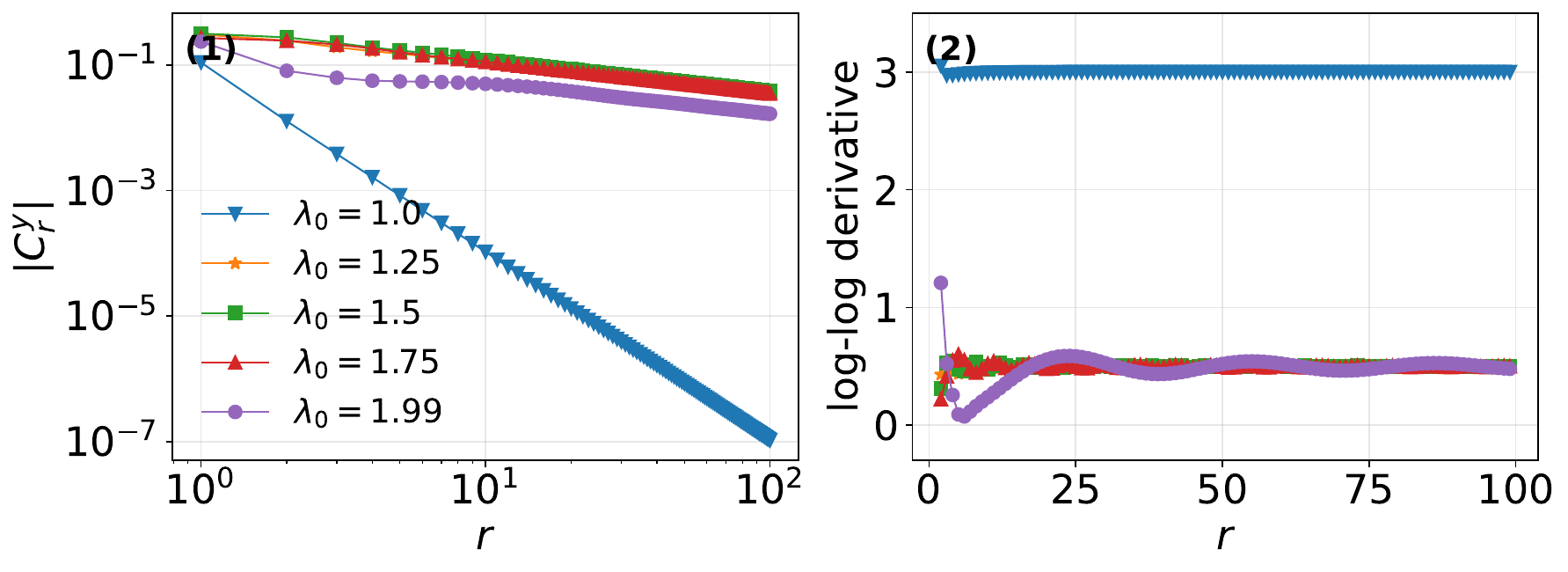} 
	\end{center}
	\vspace{-14pt}  
	\caption{The $r$-dependence of the modulus of the LR $y$-correlation function $\left|C_r^y\right|$ for different $\lambda_0$, with $\lambda=\lambda_0\exp(\mi\pi/3)$, at $\gamma=1$ in the LL phase. Panel (1) shows  the bare data presented on a log-log scale. Panel (2) shows the log-log derivative $-\dd\ln|C_r^x| / \dd \ln r$.}         
	\label{com_y_lr_ll_ga}
\end{figure}

Figure \ref{com_y_lr_ll_ga} shows that $\left|C_r^y\right|$ computed using the LR expectation value decays algebraically in the LL phase with an universal exponent of $1/2$. This is identical to the behavior of $\left|C_r^x\right|$ computed within this formalism. However, at the FM–LL phase transition, the large-$r$ behavior of $\left|C_r^y\right|$ and $\left|C_r^x\right|$ differ qualitatively; $\left|C_r^x\right|$ approaches a constant, see the orange curve in Fig.~\ref{com_x_lr_lh_ga} (a1), while $\left|C_r^y\right|$ decays as a power law with exponent 3, see the blue curves in Fig.~\ref{com_y_lr_ll_ga} (1) and (2). In Appendix~\ref{ll_ex}, we analytically derive this exponent using the Fisher-Hartwig theorem. 

\subsubsection{$C_r^{x/y}$ in the PM phase and at the LL to PM transition}
\label{sec:PM_phase}

\begin{figure}[tbp] 
	\begin{center} 
		\includegraphics[width=1\linewidth]{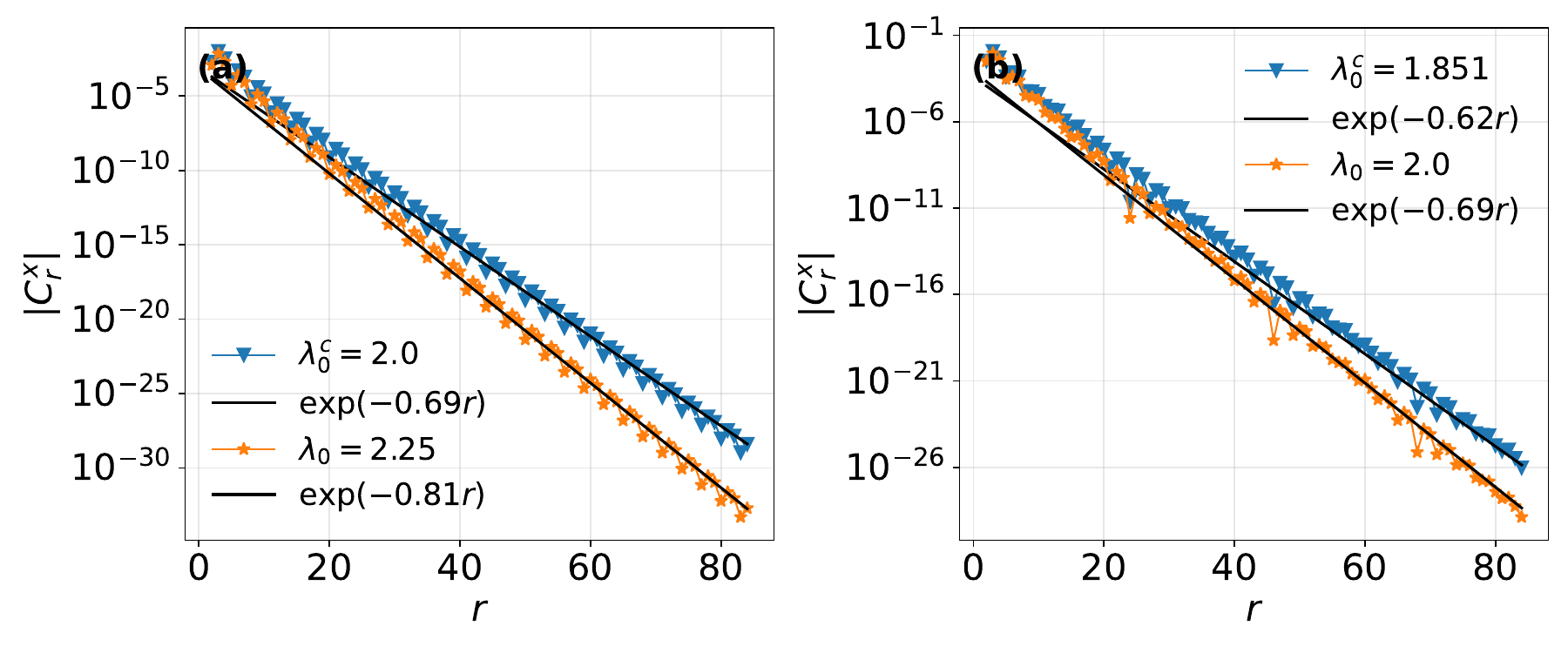} 
	\end{center}
	\caption{The $r$-dependence of the RR $x$-correlation function $|C_r^x|$ for different $\lambda_0$, with $\lambda=\lambda_0\exp(\mi\pi/3)$ (a) and $\lambda=\lambda_0\exp(\mi)$ (b), at $\gamma=1$ in the PM phase. Note  the linear-log scale.  The $\lambda_0^{\rm c}$ denotes the critical point of the LL-PM transition. The black lines show analytical results for the inverse correlation length; see Eq. \eqref{pm_ex_ga1}.}   
	\label{com_pfx_rr_pm}
\end{figure}

\begin{figure}[tbp] 
	\begin{center} 
		\includegraphics[width=1\linewidth]{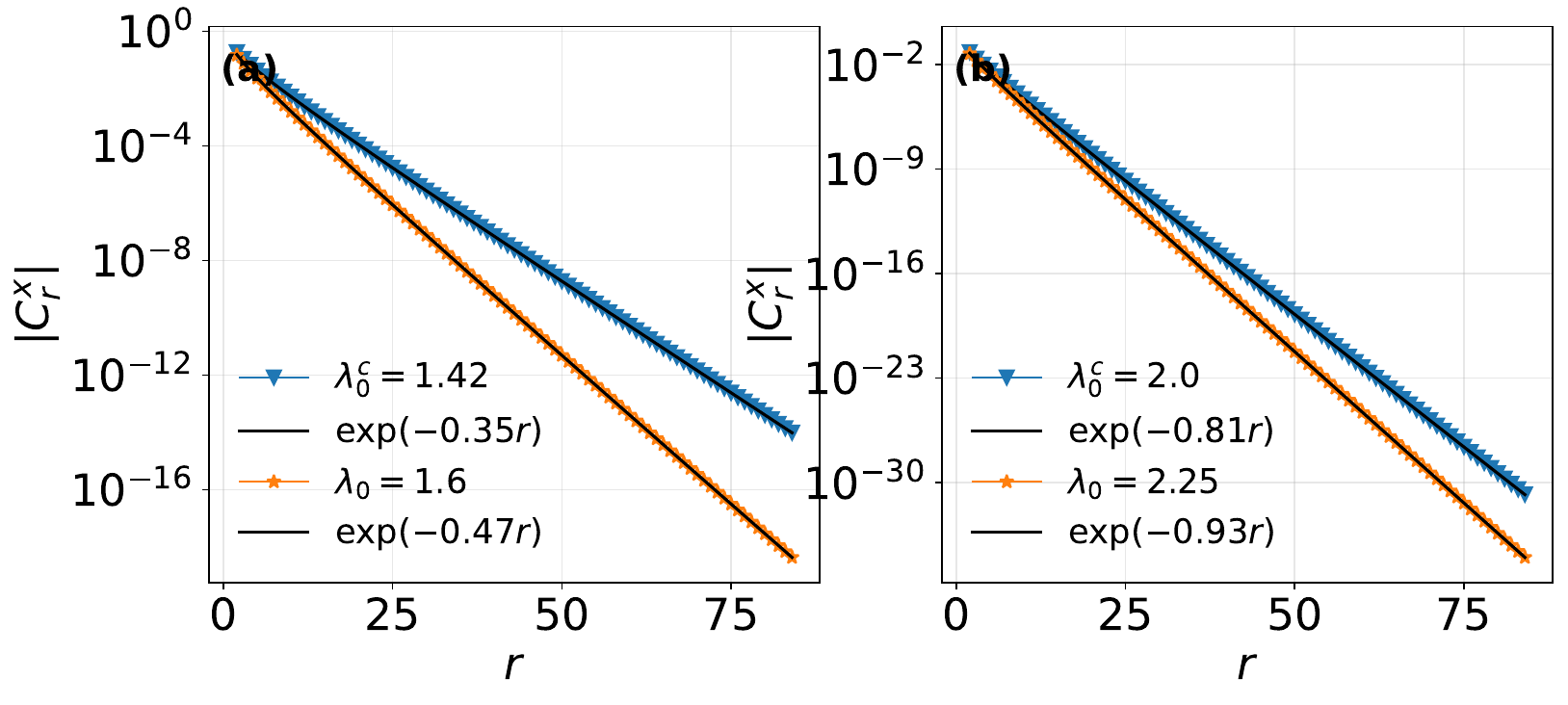} 
	\end{center}
	\caption{The $r$-dependence of the modulus of the LR $x$-correlation function $|C_r^x|$ for different $\lambda_0$, with $\lambda=\lambda_0\exp(\mi\pi/4),\gamma=1$ (a) and $\lambda=\lambda_0\exp(\mi\pi/3),\gamma=0.8$ (b) in the PM phase. Note  the linear-log scale. The black lines show analytical results for the inverse correlation length; see Eqs. \eqref{pm_ex_ga1} and \eqref{pm_ex_ga}.}  
	\label{com_pfx_lr_pm}
\end{figure}

In the PM phase and at the transition from the LL to the PM phase both the $x$- as well as the $y$-correlation functions exhibit exponential decay. Figures  \ref{com_pfx_rr_pm}  and \ref{com_pfx_lr_pm} show the RR and LR $x$-correlation functions, respectively, while the results for $C_r^y$ are deferred to Appendix~\ref{com_pfy_pm}. We here restrict ourselves and only present the bare data on a linear-log scale. This has two reasons. (1) The exponential behavior of the envelope of the data is evident and (2) for the RR expectation value leading oscillations require a more sophisticated approach when aiming at the inverse correlation length via a derivative of the numerical data. This is discussed in Appendix~\ref{com_pfy_pm}.  

Both the oscillatory and exponential behavior of the correlation functions can be analytically understood from the pole structure of the energy spectrum in the complex momentum plane. In particular, the real part of the pole determines the oscillation frequency, while the imaginary part determines the inverse correlation length and therefore governs the exponential decay. For $\gamma^2=1$ one finds from the dispersion relation Eq.~\eqref{e} that the energy pole satisfies
\begin{equation}
e^{\mi k} = \lambda^{-1} = |\lambda|^{-1} e^{-\mi\phi},
\end{equation}
leading to the following asymptotic form of the correlation functions
\begin{equation}
|C_r^{x/y}| \sim e^{\mi kr} \sim |\cos(\phi r)|\, e^{-r \ln|\lambda|},\label{pm_ex_ga1}
\end{equation}
thereby indicating an oscillation period of $\pi/\phi$ and a correlation length $\xi= 1/\ln|\lambda|$. When $\phi/\pi$ is a rational number, the pattern is commensurate with respect to the lattice, as illustrated in Figs.~\ref{com_pfx_rr_pm} (a), where the correlations exhibit a periodic exponential decay within a uniform envelope. In contrast, when $\phi/\pi$ is irrational, the decay pattern is incommensurate, as shown in Fig.~\ref{com_pfx_rr_pm} (b), where no well-defined period can be identified. When $\gamma^2\neq1$, the oscillations become even less regular and no clear periodicity is observed.

As exemplified in Fig.~\ref{com_pfx_lr_pm} the absolute value of the LR $x$-correlation function does not show any oscillations; they only show up in the individual real and imaginary components. Despite this differences, the exponential decay rate remains identical for both RR and LR as well as $x$- and $y$-correlation functions. 

By analyzing the poles of the energy spectrum one can show that the inverse correlation length $\xi^{-1}$ is determined by the dominant one among the four poles
\begin{equation}
\mathcal{Z} = \left\{ \frac{\lambda \pm \sqrt{\lambda^2+\gamma^2-1}}{1 \pm \gamma}\right\}.
\end{equation}
Since each pole contributes an exponential factor of the form $z^r$ to the correlation function, the long-distance behavior is controlled by the pole whose magnitude is smaller than but closest to unity. Therefore, the inverse correlation length $\xi^{-1}$ is governed by the pole $z_* \in \mathcal{Z}$ satisfying
\begin{equation}
\xi^{-1} = -\ln |z_*|, \quad z_* = \text{max} { |z| : z \in \mathcal{Z}, |z| < 1 }.\label{pm_ex_ga}
\end{equation}
We thus obtain an analytical expression for $\xi$ also for $\gamma^2  \neq 1$; compare to Eq.~\eqref{pm_ex_ga1}. In Figs.~\ref{com_pfx_rr_pm} and \ref{com_pfx_lr_pm}, we superimpose these analytical predictions Eqs.~\eqref{pm_ex_ga1} and \eqref{pm_ex_ga} onto the numerical results, showing excellent agreement.

\subsubsection{RR versus LR}

To provide an overview of our results, we summarize the asymptotic behavior of the correlation functions in Table~\ref{tab1}. In this we also compare the behavior obtained from the RR and LR expectation value.  

As already emphasized when investigating open quantum systems employing the formalism of standard quantum mechanics and thus use RR expectation values is more reasonable on general grounds and avoids any unphysical results \cite{Meden2023} such as, e.g., complex expectation values; see the complex magnetization or the complex correlation functions above. 

For the non-Hermitian complex-$\lambda$ model it was mentioned that, when employing biorthogonal quantum mechanics with its LR expectation values, the critical properties of $C_r^{x/y}$ are closer to those of the Hermitian XY model with magnetic field \cite{Wang2025}. Our results confirm this; compare Tables \ref{tab_her} and \ref{tab1}. Most prominently, the power-law decay in the LL phase has exponent 1/2 independent of the parameters and $C_r^x$ approaches a non-vanishing constant while $C_r^y$ goes to zero exponentially in the FM phase with $\gamma>0$. This is related to the observation that the analytical expressions for the correlation functions of the minimal energy state evaluated as LR expectation values are closer in form to those obtained for the Hermitian model; one can speak of an analytic continuation of the Hermitian expressions. Note however, that this cannot be considered as an argument in favor of the use of the biorthogonal formalism. We are convinced that if the properties investigated here are experimentally accessible at all, a properly executed experiment would show the richer and more interesting critical properties derived using standard quantum mechanics with RR expectation values.   

\begin{table}[tbp]
	\centering
	\begin{tabular}{ccccccc}
		\toprule\toprule
		\multicolumn{2}{c}{$r \to \infty$} & FM & FM-LL & LL & LL-PM & PM \\ \midrule
		\multirow{2}{*}{$\left|C_r^x\right|$} & RR & const.& const. & $r^{-\alpha}$  & \multicolumn{2}{c}{\multirow{4}{*}{$e^{-r/\xi}$}} \\
		& LR & const. & const. & $r^{-1/2}$ & \multicolumn{2}{c}{} \\ \cmidrule(r){1-5} 
		\multirow{2}{*}{$\left|C_r^y\right|$} & RR & const. & const. & $r^{-\alpha}$  & \multicolumn{2}{c}{} \\
		& LR & $e^{-2r/\xi}$ & $r^{-3}$ & $r^{-1/2}$ & \multicolumn{2}{c}{} \\ \bottomrule\bottomrule
	\end{tabular}
		\caption{Asymptotic behavior of the correlation functions in the complex-$\lambda$ XY model at $\gamma>0$ for the minimal energy state. Here, $0 < \alpha < 1/2$, and the inverse correlation length is $\xi^{-1} = \left|\ln \left|\lambda\right|\right|$ at $\gamma = 1$, and $\xi^{-1} = -\ln |z_*|$ for $\gamma \neq 1$. The results for $\gamma < 0$ are equivalent to those for $\gamma > 0$ when interchanging the $x$ and $y$ components.}\label{tab1}
\end{table}

\subsubsection{Correlation functions in the steady state}

\begin{figure*}[htb]  
	\begin{center} 
		\includegraphics[width=0.99\linewidth]{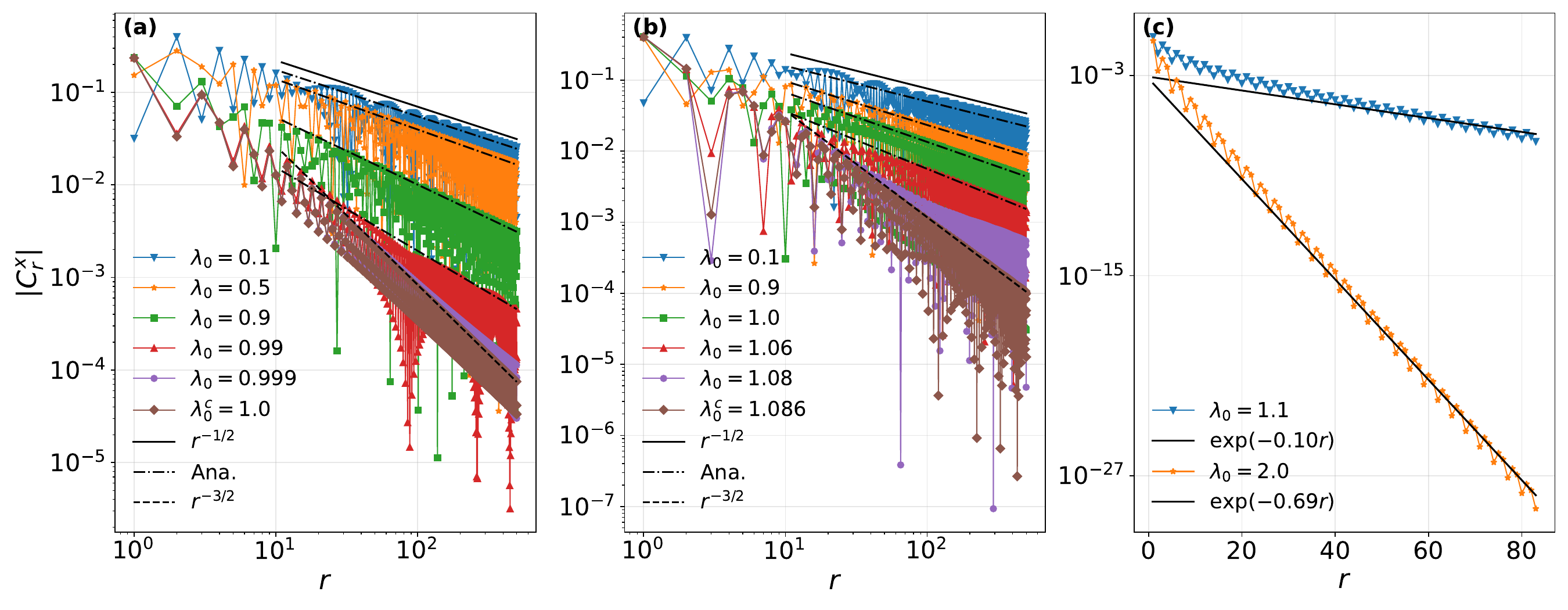}\end{center} 
	\caption{The $r$-dependence of the RR $x$-correlation function $C_r^x$ for different $\lambda_0$, with $\lambda=\lambda_0\exp(\mi\pi/3),\gamma=1$ (a), $\lambda=\lambda_0\exp(\mi\pi/4),\gamma=1.2$ (b) and $\lambda=\lambda_0\exp(\mi\pi/3),\gamma=1$ (c) in the steady state. Panels (a) and (b) show data from the LL phase and at the LL-PM transition; note the log-log scale. The black lines indicate a power law with the exponent given by Eq.~\eqref{cho2LL}. Panel (c) shows data from the PM phase on a linear-log scale. The black lines show exponential behavior with the inverse correlation length given by Eq.~\eqref{pm_ex_ga}.}         
	\label{cho2}
\end{figure*}

As a last step in our study of the complex-$\lambda$ model we briefly discuss results for $C_r^{x}$ in the steady state. As shown above in this we only find a LL and a PM phase. 
 
Figure \ref{cho2} shows the $r$-dependence of the RR $x$-correlation function for different parameter sets. Panel (a) and (b) show data in the LL phase and at the LL-PM transition on a log-log scale. It is apparent that the envelope shows power-law scaling. The oscillatory fine structure prevents us from taking the log-log derivative. The exponent varies between $1/2$ and $3/2$. Extremely close to the LL-PM phase transition, the exponent changes rapidly; note the small $\lambda_0$-steps close to $\lambda_0^{\rm c}$ in panels (a) and (b). 

Considering the block case of the Fisher-Hartwig theorem \cite{Basor2025} the exact analytical expression for the exponent remains elusive as the jump discontinuities associated with the steady state do not satisfy the required assumption. Nevertheless, away from the immediate vicinity of the LL-PM transition, an approximate result can be derived. The corresponding power-law exponent is $(\theta_0/\pi)^2+(\theta_0/\pi+1)^2$  with
\begin{align}
&\cos \theta_0=-\frac{\I (\lambda)}{m}, \quad \sin \theta_0=-\frac{e}{m},\no\\
&m=\gamma\sqrt{1-\R^2 (\lambda)}, \quad e=\sqrt{m^2-\I^2 (\lambda)}, \label{cho2LL}
\end{align}
whose range is $[1/2,1]$.
The numerical data of Fig.~\ref{cho2}  (a) and (b) are overlaid with a power law with the exponent given by this analytical expression (black dashed-dotted lines). The agreement is convincing.  

Panel (c) of Fig.~\ref{cho2} shows data in the PM phase. In this the correlation function decays exponentially with the inverse correlation length given by Eq.~\eqref{pm_ex_ga}; an exponential decay with this is displayed as the black lines.

\section{Imaginary‑$\gamma$ model}

\subsection{Energy and magnetization}

In this section, we consider the model with purely imaginary anisotropy with the Hamiltonian ($\lambda \in {\mathbb R})$ \cite{Zhang2013a,Zhang2013b,Miao2024,Zhang2025}
\begin{equation}
H = -\frac{J}{2}\sum_{l=1}^N \left( \frac{1+\mi\gamma}{2} \, \sigma_l^x \sigma_{l+1}^x 
+ \frac{1-\mi\gamma}{2} \, \sigma_l^y \sigma_{l+1}^y \right)
- \frac{\lambda}{2} \sum_{l=1}^N \sigma_l^z. \label{Hamn}
\end{equation}
The model possesses an anti-unitary symmetry given by the combined action of a unitary spin-rotation by $\pi/2$ around the $z$-axis 
\begin{align}
\mathcal{R} = \exp\left[ -\mi(\pi/4) \sum_{l=1}^N \sigma_l^z \right]
\label{rotation}
\end{align}
and the anti-linear complex conjugation ${\mathcal K}$ acting as $\mathcal{K}\mi \mathcal{K}^{-1} = -\mi$. As can easily be checked $[\mathcal {RK},H]=0$ \cite{footnote3}.   

In full analogy to well-studied parity-time symmetric non-Hermitian models \cite{Benderbook,Meden2023}, $\mathcal{RK}$-symmetry implies that all eigenvalues of the Hamiltonian are either real ($\mathcal{RK}$-symmetric phase) or that the eigenvalues are partly real and partly complex conjugate to each other ($\mathcal{RK}$-broken phase). As long as an eigenstate of the Hamiltonian is also an eigenstate of $\mathcal{RK}$ its corresponding eigenvalue is real. The model thus allows to study the interesting interplay between the $\mathcal{RK}$-symmetry breaking transition and magnetic phase transitions. 

\begin{figure}[t] 
	\begin{center} 
		\includegraphics[width=0.9\linewidth]{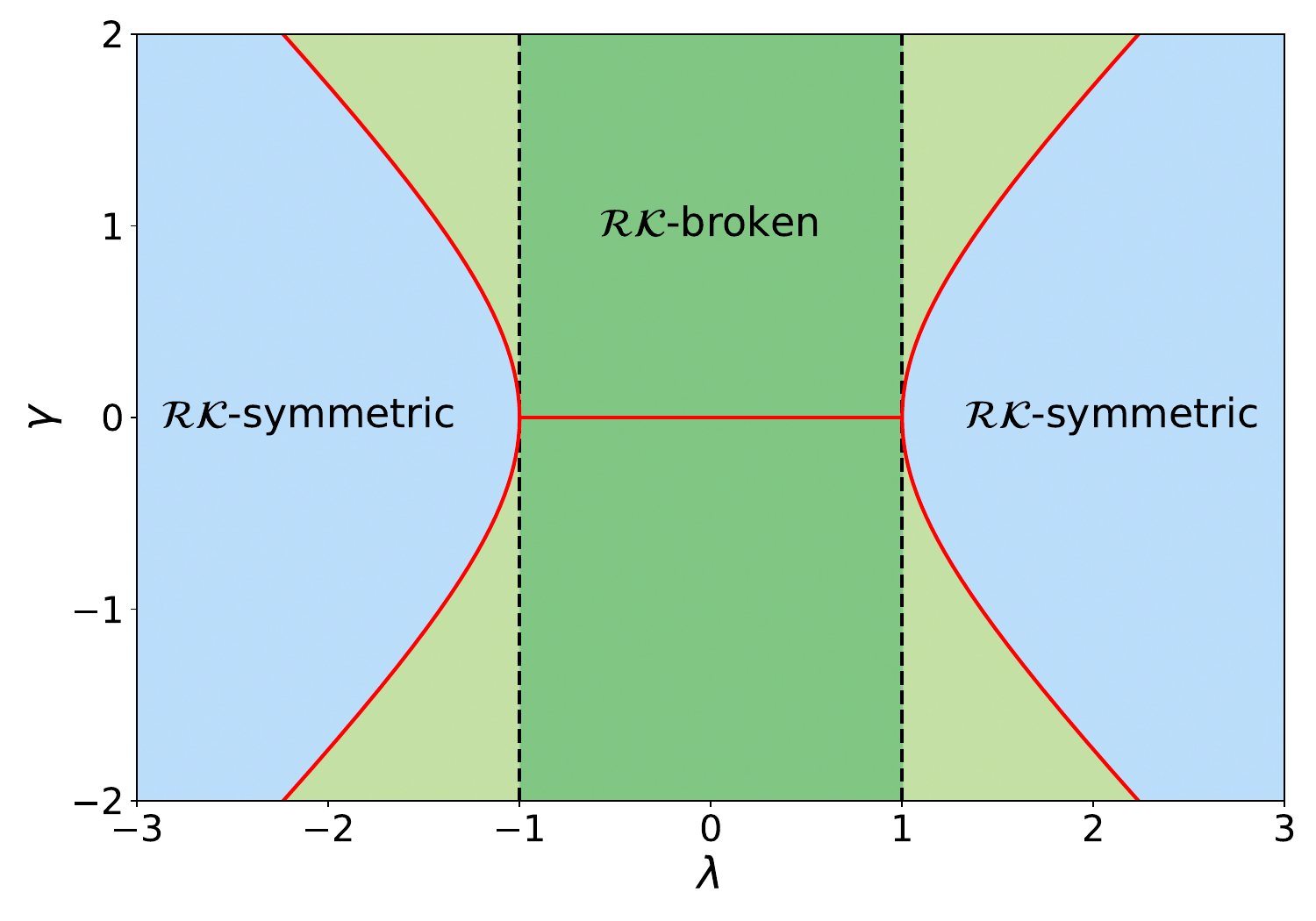} 
	\end{center}
	\vspace{-14pt}  
	\caption{The phase diagram of the imaginary‑$\gamma$ XY model. The red lines denote the  $\mathcal{RK}$ symmetry-breaking transition (hyperbolas) and the Hermitian limit ($\gamma=0$). The black dashed line at $|\lambda|=1$ marks the boundary between distinct magnetic properties. }  
	\label{pd_im}
\end{figure}

For the imaginary-$\gamma$ model the quasi-particle energies $E_k$ Eq.~\eqref{e} are given by
\begin{equation}
E_k=\pm \sqrt{(\lambda - \cos k)^2-(\gamma \sin k)^2},\label{en}
\end{equation}
which becomes purely imaginary for momenta $k$ with $(\lambda - \cos k)^2 < (\gamma \sin k)^2$. An imaginary quasi-particle energy implies that eigenenergies of the Hamiltonian Eq.~\eqref{Hamn} become complex. Accordingly, the phase boundary between the $\mathcal{RK}$-symmetric phase with entirely real spectrum and the $\mathcal{RK}$-broken phase with complex eigenvalues is determined by 
\begin{equation}
E_k=0, \quad \frac{\partial E_k}{\partial k}=0
\end{equation}
for one $k \in [0,\pi]$. This yields the phase boundaries
\begin{eqnarray}
\mathcal{RK}\text{-symmetric phase: } &&\lambda^2-\gamma^2>1 \text{ or }  \gamma=0,\no\\
\mathcal{RK}\text{-broken phase: } &&\lambda^2-\gamma^2<1 \text{ and }  \gamma\neq0.
\label{rt}
\end{eqnarray}
The corresponding phase diagram is shown in Fig.~\ref{pd_im}. As discussed below also the magnetic properties change across the (red) lines of $\mathcal{RK}$-symmetry breaking. The black dashed lines at $|\lambda|=1$ mark  additional boundaries between parameter regimes with distinct magnetic properties; see below.

For the quasi-particle energies in Eq.~\eqref{en} we again have to select one of the signs for each mode $k$. As for the complex-$\lambda$ model this fixes the physical properties of the quasi-particle vacuum state we focus on. This eigenstate has energy
\begin{equation}
E_{\mathrm{Vac}}=-{\sum_{k}}^{'}\left( \pm \sqrt{(\lambda - \cos k)^2-(\gamma \sin k)^2}+\cos k \right).
\label{vac_energy}
\end{equation}
For modes with real $E_k$ in Eq.~\eqref{en} we take the plus sign such that the vacuum state is the eigenstate with the lowest real part of the energy. For modes for which the quasi-particle energy is purely imaginary we take the minus sign such that the vacuum state in addition has the largest imaginary part of all eigenstates. The state we take is thus the natural combination of the lowest-energy and the steady state we investigated for the complex $\lambda$-model. 

Consider the $\mathcal{RK}$-broken phase. In this phase at least one of the mode energies in Eq.~\eqref{en} contributing to the vacuum state energy Eq.~\eqref{vac_energy} is purely imaginary. The partner eigenstate to the vacuum state which, according to $\mathcal{RK}$-symmetry, has the energy $E_{\mathrm{Vac}}^\ast$ is the one in which all the $k$ and $-k$ quasi-particle states corresponding to purely imaginary $E_k$ are occupied; see Eq.~\eqref{fh}.     

\begin{figure}[tb] 
	\begin{center} 
		\includegraphics[width=0.9\linewidth]{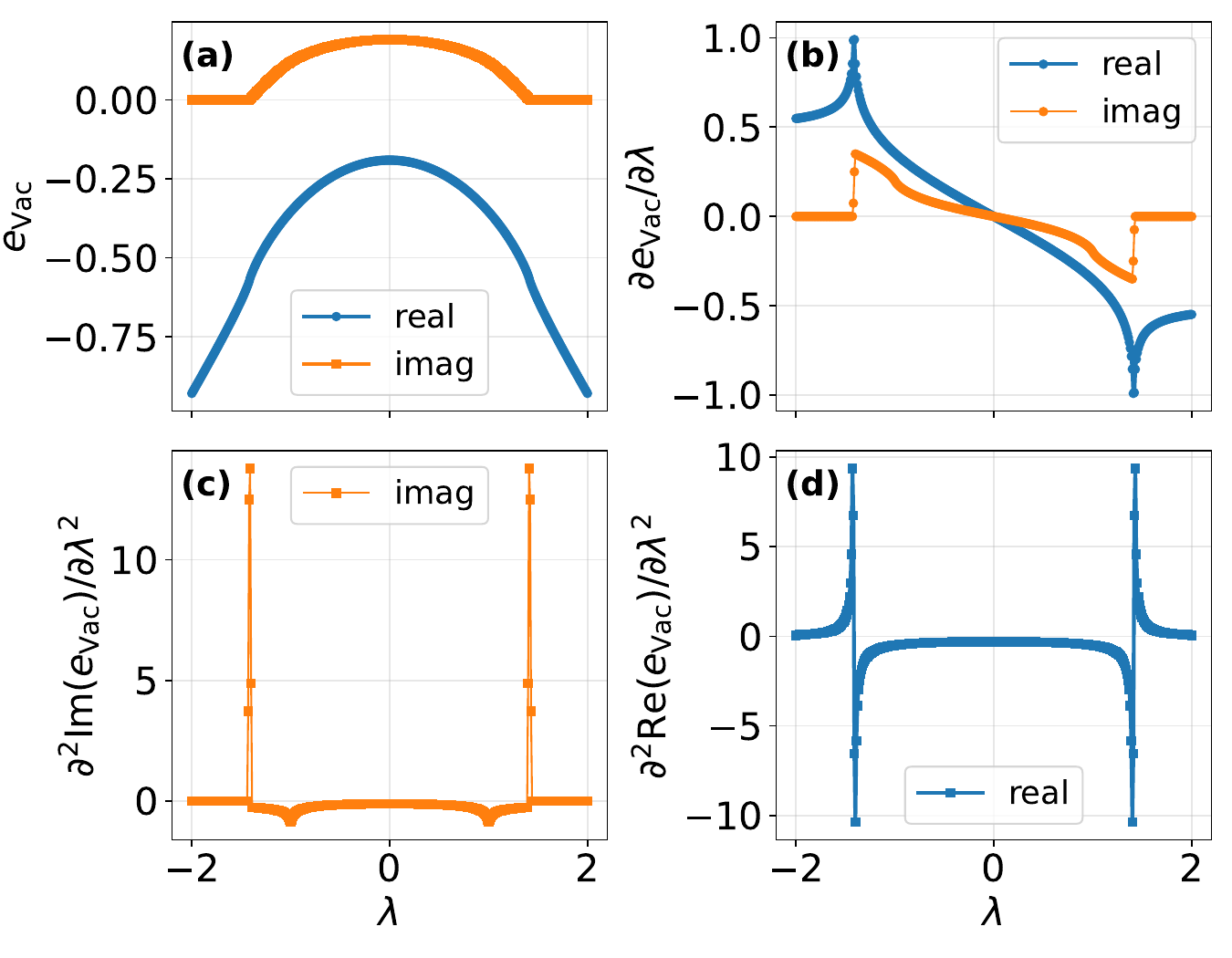} 
	\end{center}
	\vspace{-14pt}  
	\caption{The vacuum state energy density (a) as a function of $\lambda$ for $\gamma=1$. Panel (b) shows its first derivative with respect to $\lambda$. Panels (c) and (d)  show the second derivatives of the imaginary and real parts, respectively.} 
	\label{im_gs_energy}
\end{figure}

In Fig.~\ref{im_gs_energy}, we present the vacuum state energy density $e_{\mathrm{Vac}}$, together with its first and second derivatives as a function of $\lambda$ for $\gamma=1$. As shown in Fig.~\ref{im_gs_energy} (a), the energy density itself remains continuous throughout the parameter space. However, its first derivative of both the real and the imaginary part in Fig.~\ref{im_gs_energy} (b) exhibit clear non-analyticities at the $\mathcal{RK}$ symmetry-breaking transition points defined by Eq.~\eqref{rt}. For the second derivatives, the imaginary and real parts in Fig.~\ref{im_gs_energy} (c) and (d) similarly show non-analytic behavior at the $\mathcal{RK}$ transition. Notably,  Fig.~\ref{im_gs_energy} (c) displaying the second derivative of the imaginary part reveals additional features at $\lambda = \pm 1$, indicating the presence of another transition (already indicated in Fig.~\ref{pd_im}).

\begin{figure*}[tb] 
	\begin{center} 
		\includegraphics[width=0.96\linewidth]{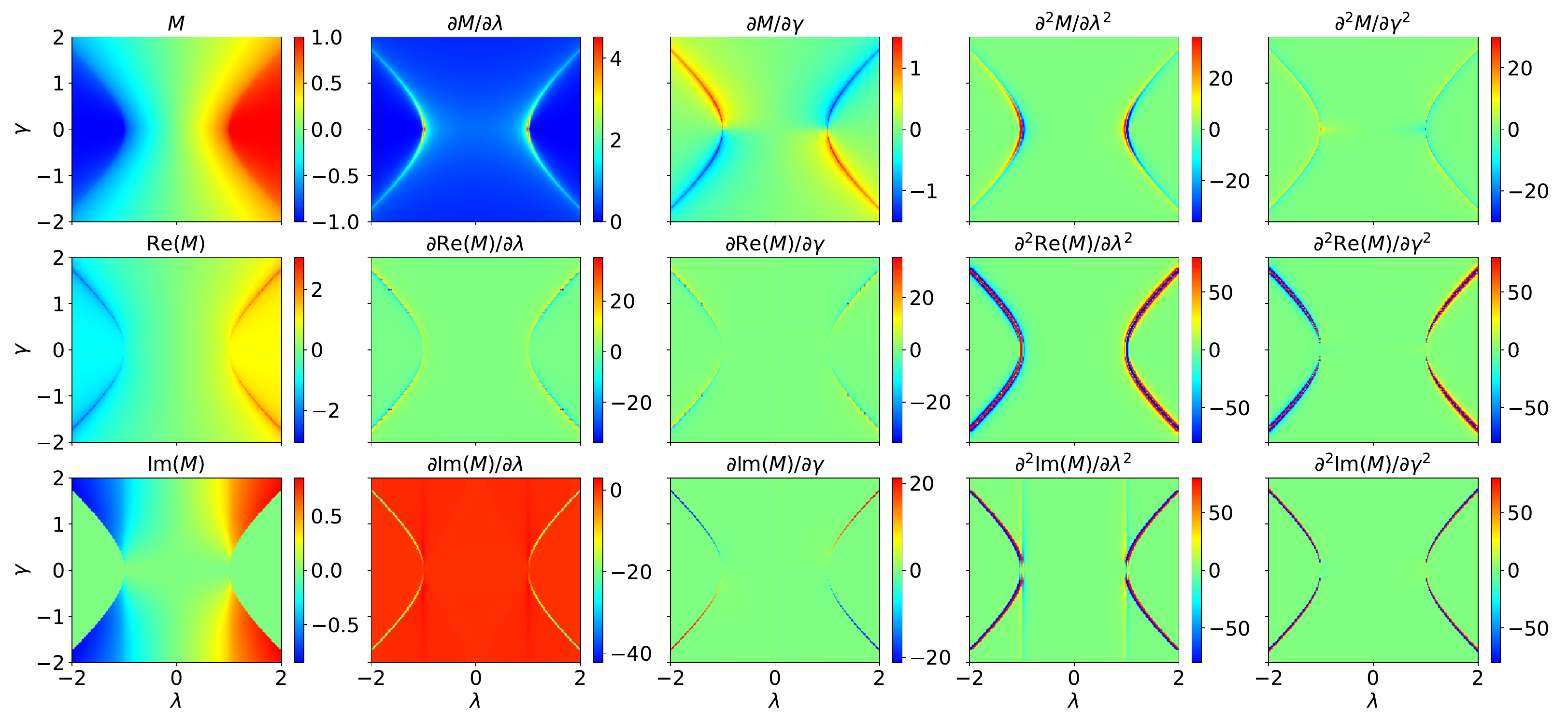} 
	\end{center}
	\vspace{-14pt}  
	\caption{The magnetization of the vacuum state as a function of $\lambda$ and $\gamma$ and its first and second derivative with respect to the two parameters. From top to bottom: the RR magnetization, the real and imaginary parts of the LR magnetization.}         
	\label{im_mag}
\end{figure*}

The RR and LR magnetization is given by Eqs.~\eqref{rr_m} and \eqref{lr_m}, where the Bogoliubov coefficients are obtained from Eq.~\eqref{uv} with $\gamma \to \mi \gamma$. In Fig.~\ref{im_mag} we show the magnetization and its derivatives as a function of $\lambda$ and $\gamma$. In the LR case, the magnetization becomes ill-defined at the critical line $\lambda=\sqrt{1+\gamma^2}$ due to the presence of a non-integrable singularity in the integrand; accordingly the numerical results are shown only in the vicinity of this line. All three panels clearly capture the $\mathcal{RK}$ symmetry-breaking phase transition. In addition, the derivatives of the imaginary part of the LR magnetization with respect to $\lambda$ exhibits a feature at $|\lambda| = 1$ as was already found in the energy density.

\subsection{Correlation functions}

The expressions for the RR and LR correlation functions $C_r^{x/y}$ derived in Sect.~\ref{subsec:correl} can also be used for the imaginary-$\gamma$ model and allow us to numerically compute those as well as to gain analytical insights. Based on the asymptotic behavior of $C_r^{x/y}$ we can distinguish the magnetic phases which overlay the $\mathcal{RK}$-symmetry phases and, as we will see, imprint fine structure on these.   

\subsubsection{The $\mathcal{RK}$-broken phase and the transition lines}

\begin{figure*}[tb] 
	\begin{center} 
		\includegraphics[width=1\linewidth]{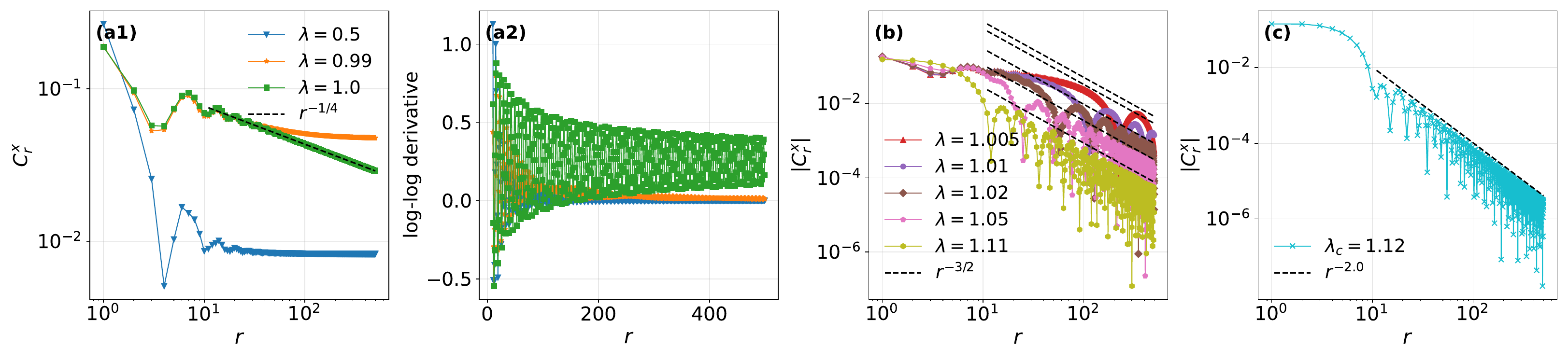} 
	\end{center}
	\vspace{-14pt}  
	\caption{The vacuum state RR $x$-correlation functions $|C_r^x|$ as a function of $r$ for $\gamma = 0.5$. Panel (a2)  shows the corresponding power-law exponents extracted from the data of (a1) by a log-log derivative. The panels (a1), (b), and (c) correspond to $\lambda \leq 1$, $1 < \lambda < \lambda_c = \sqrt{1+\gamma^2}$, and $\lambda = \lambda_c$, respectively, with the data in these three columns plotted on a log–log scale. The dashed black lines in (a1), (b), and (c) show power laws with the indicated exponents.} 
	\label{rr_lh}
\end{figure*}

Figure \ref{rr_lh} shows the $r$-dependence of the $x$-direction correlation function using the RR expectation value for $\gamma =0.5$ and $-\lambda_c \leq \lambda \leq \lambda_c= \sqrt{1+\gamma^2}$, i.e., in the $\mathcal{RK}$-broken phase and at the transition. Note the log-log scale of panels (a1), (b) and (c). For $-1<\lambda<1$ the correlation function approaches a constant at large $r$ characteristic for a FM phase; see Fig.~\ref{rr_lh} (a1) and (a2) for the log-log derivative. At $|\lambda|=1$, the envelope of $C_r^x$ shows power-law scaling with exponent $1/4$, however, overlaid by subleading oscillations; see the green lines in Figs.~\ref{rr_lh} (a1) and (a2). For $1 < |\lambda| < \lambda_c$ the envelope of the $x$-correlation function decays as $r^{-3/2}$, see  Fig.~\ref{rr_lh} (b), and at $\lambda_c$ as $r^{-2}$, see Fig.~\ref{rr_lh} (c). For the latter two cases oscillations are already very prominent in the bare data such that we refrain from presenting results for the log-log derivatives. We note in passing that the oscillation frequency can be understood analytically from the the poles of the energy spectrum. However, the current formulation of the Fisher-Hartwig theorem does not allow to obtain analytical results for the scaling exponents. 

The numerical results show that in the light green part of the phase diagram Fig.~\ref{pd_im} the vacuum state shows quasi-long-range magnetic order. This is reminiscent of the LL phase of the Hermitian model although with a different critical exponent.   

For $\gamma=0$ we recover the isotropic XX model with magnetic field and (with our choice of signs in $E_k$) the vacuum state becomes the ground state of the Hermitian Hamiltonian. For $|\lambda|<1$ the model is in its LL phase; see Sect.~\ref{sec_Herm}.  

The $y$-direction correlation function shows the same characteristics and is not presented here. Some symmetry relations for $C_r^x$ and $C_r^y$ are discussed in Appendix \ref{app_sym}.

\begin{figure*}[tb] 
	\begin{center} 
		\includegraphics[width=0.9\linewidth]{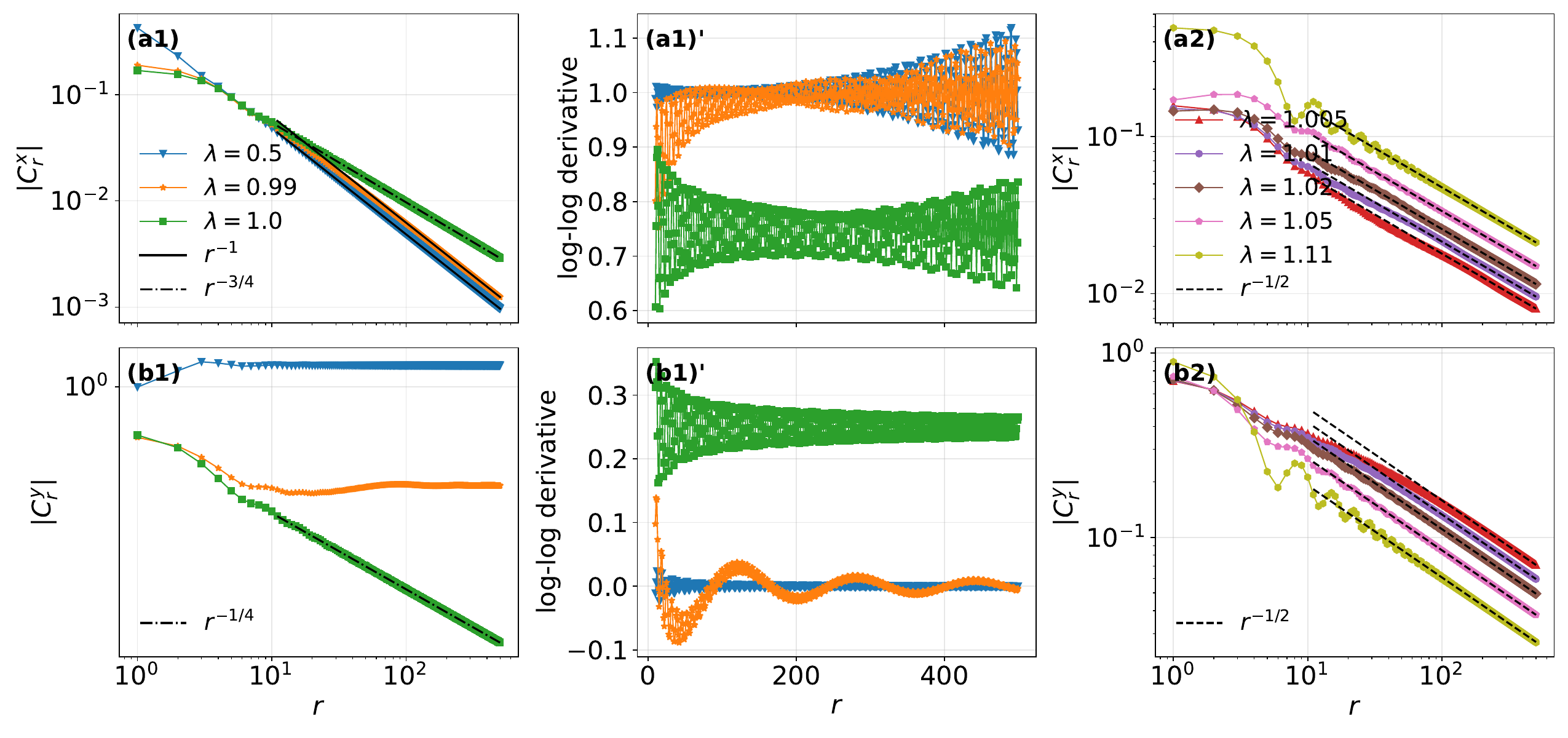} 
	\end{center}
	\vspace{-14pt}  
	\caption{The vacuum state LR $x$- (upper row) and $y$- (lower row) correlation functions as a function of $r$ at $\gamma = 0.5$. Panel (a1)' and (b1)' show the corresponding power-law exponents extracted from the data of (a1) and (b1) by a log-log derivative. The panels (a1)(b1) and (a2)(b2) correspond to $\lambda \leq 1$ and $1 < \lambda < \lambda_c = \sqrt{1+\gamma^2}$, respectively, with the data in these two columns plotted on a log–log scale. The dashed black lines in (a1)(b1) and (a2)(b2) show power laws with the indicated exponents.}         
	\label{lr_lh}
\end{figure*}

We next discuss the asymptotic $r$-behavior of $|C_r^x|$ and $|C_r^y|$ obtained by taking the LR expectation value with the vacuum state. For $\gamma=0.5$ it is shown in Fig.~\ref{lr_lh} on a log-log scale. For $\lambda<1$ the $y$-correlation function shown in panel (b1) saturates at a finite value while $C_r^x$ of (a1) goes to zero. One would identify this with a FM$_y$ phase. However, the decay to zero of $|C_r^x|$ is not exponential as in the FM$_y$ phase of the Hermitian model, but rather power-law like with exponent 1; see panel (a1)' showing the log-log derivative.       

At the critical lines $|\lambda| = 1$, $|C_r^x|$ and $|C_r^y|$ both go to zero with exponents $3/4$ and $1/4$, respectively. This is shown in Fig.~\ref{lr_lh} (a1), (a1)', (b1), and (b1)'. In the parameter regime $1 < \lambda < \lambda_c$, as shown in (a2) and (b2), $|C_r^x|$ and $|C_r^y|$ show the same critical scaling with exponent $1/2$. This behavior is again reminiscent of a LL phase this time even with the same exponent as in the Hermitian model. The power-law exponents of the LR correlation functions are derived analytically in Appendix \ref{fh_imag}. At $\lambda = \lambda_c$, the expression for the LR correlation function in Eq.~\eqref{elr} exhibits a non-integrable singularity of the form $|k - k_0|^{-1}$. Therefore, the LR correlation function becomes ill-defined at the boundary between the $\mathcal{RK}$-broken and $\mathcal{RK}$-symmetric phase. 

It can be also observed from Figs.~\ref{rr_lh} and \ref{lr_lh} that oscillations exist either in leading or subleading order of the correlation functions. A more detailed analysis of this is deferred to Appendix \ref{im-oscis}.

\subsubsection{The $\mathcal{RK}$-symmetric phase}

\begin{figure}[tb] 
	\begin{center} 
		\includegraphics[width=1\linewidth]{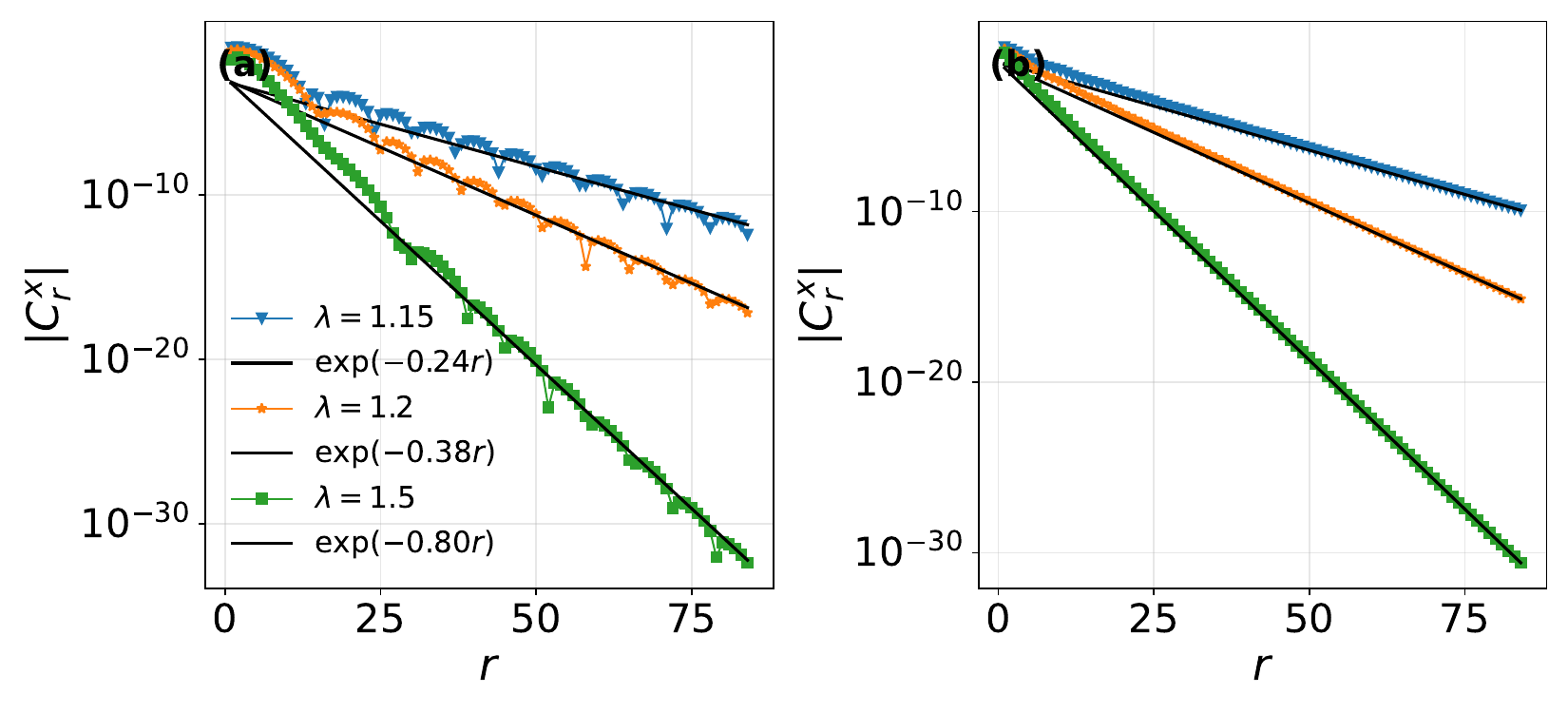} 
	\end{center}
	\vspace{-14pt}  
	\caption{The $r$-dependence of the correlation functions $|C_r^x|$ at $\gamma=0.5$ for the RR expectation value (a) and the LR expectation value (b) in the $\mathcal{RK}$-symmetric phase.}         
	\label{pm}
\end{figure}

In the $\mathcal{RK}$-symmetric phase, the real energy spectrum is gapped, and the RR and LR correlation functions $C_r^x$ and $C_r^y$ exhibit exponential decay; the system is in a PM phase. Figure \ref{pm} shows the $r$-dependence of $|C_r^x|$ computed taking the RR as well as the LR expectation value on a linear-log scale for $\lambda > \lambda_c$. For both types of expectation values the inverse correlation length $\xi^{-1}$ is analytically determined by the leading contribution among the following four solutions
\begin{equation}
\mathcal{Z} = \left\{ \frac{\left(\lambda \pm \sqrt{\lambda^2-\gamma^2-1}\right)\left(1\pm \mi\gamma\right)}{1 + \gamma^2}\right\}.\label{pm_br}
\end{equation}
In particular, the decay is controlled by the pole $z_*$ in the set $\mathcal{Z}$ whose modulus is strictly less than, but closest to unity, i.e., $\xi^{-1} = -\ln |z_*|$. Exponentially decaying functions with this result for the inverse correlation length are shown in Fig.~\ref{pm} as black lines; the agreement with the numerical data is excellent. Equation \eqref{pm_br} further indicates that the frequency of the oscillations, visible in the RR expectation value in the $\mathcal{RK}$-symmetric phase, is given by $\arctan(\gamma)$, which is independent of $\lambda$. This frequency also coincides with that obtained from $k_0^{\pm}$ in Appendix Eq.~\eqref{zeros} at the transition point $\lambda = \lambda_c$. 

\subsubsection{RR versus LR}

As for the Hermitian XY model with magnetic field and the complex-$\lambda$ model we summarize our findings for the asymptotic $r$-dependence of the correlation functions in Table \ref{tab2}. This also allows for a direct comparison of the results for the RR and the LR correlation functions. 

We observe that the LR results are again closer to the ones of the Hermitian model while characteristic differences exist, such as, e.g., the power-law instead of exponential decay of $C_r^x$ in a FM$_y$-like phase in which $C_r^y$ approaches a constant. We reemphasize that this closeness cannot be regarded as a argument in favor of the formalism of biorthogonal quantum mechanics with its LR expectation values \cite{Wang2025}. A hint that this formalism should not be used when studying open quantum systems specific to the imaginary-$\gamma$ model is the ill-definedness of the correlation function on the critical line of $\mathcal{RK}$-symmetry breaking.

\begin{table}[tbp]
	\centering
	\begin{tabular}{ccccccc}
		\toprule\toprule
		\multicolumn{2}{c}{} & \multicolumn{3}{c}{$\mathcal{RK}$-broken} &   & $\mathcal{RK}$-symmetric \\ \midrule
		\multicolumn{2}{c}{$r \to \infty$} & FM & FM-LL & LL & LL-PM & PM \\ \midrule
		\multirow{2}{*}{$|C_r^x|$} & RR & const.& $r^{-1/4}$ & $r^{-3/2}$  & $r^{-2}$&\multirow{4}{*}{$e^{-r/\xi}$} \\
		& LR &$r^{-1}$ & $r^{-3/4}$ & $r^{-1/2}$ & None &  \\
		 \cmidrule(r){1-6} 
		 \multirow{2}{*}{$|C_r^y|$} & RR & const.& $r^{-1/4}$& $r^{-3/2}$  &$r^{-2}$& \\
		& LR & const. & $r^{-1/4}$& $r^{-1/2}$ & None & \\
		 \bottomrule\bottomrule
	\end{tabular}
	\caption{Asymptotic behavior of the correlation functions in the imaginary-$\gamma$ XY model at $\gamma>0$ for the vacuum state. Here, the inverse correlation length is $\xi^{-1} = -\ln |z_*|$. }\label{tab2}
\end{table}

\section{Conclusion}

In this work we analyzed the quantum critical properties of two non‑Hermitian extensions of the one‑dimensional XY spin‑1/2 chain with magnetic field, focusing on a model with complex magnetic field $\lambda$ and a model with purely imaginary anisotropy $\gamma$. The motivation was threefold. Foremost, we wanted to provide a comprehensive picture of the rich emergent quantum critical properties which can occur in such non-Hermitian many-body systems. Prior to our study, the state of the field was unsatisfactory, partly due to a lack of analytical results and an improper interpretation of numerical data. In addition, we were aiming to clarify two conceptual ambiguities that arise in non‑Hermitian many‑body systems: which formalism should be used to define expectation values and which eigenstate should play the role of the ground state when the spectrum is complex.

To address the first issue we systematically compared observables and correlation functions computed within standard quantum mechanics (right–right expectation values) and within biorthogonal quantum mechanics (left–right expectation values). While both approaches sometimes yield qualitatively similar results, we demonstrated that the biorthogonal formulation can produce complex-valued observables or even ill‑defined correlators, whereas the standard formulation always leads to physically meaningful quantities. These findings provide further evidence that the standard quantum mechanical definition of expectation values is the physically appropriate one for (at least in priciple) experimentally realizable non‑Hermitian open quantum systems.

To address the second issue we investigated several natural eigenstates: the state with minimal real part of the energy (“minimal energy state”) as well as the state with maximal imaginary part (“steady state”) for the complex-$\lambda$ model and a state which combines both features for the imaginary-$\gamma$ model. By computing the energy density, magnetization, and long‑distance asymptotics of spin correlation functions we showed that the resulting phase diagrams and critical behavior depend strongly on which of these states is considered. For the complex‑$\lambda$ model the minimal‑energy state exhibits ferromagnetic (FM), Luttinger‑liquid (LL), and paramagnetic (PM) regimes, whereas the steady state displays only LL and PM phases. For the imaginary‑$\gamma$ model we identified a rich structure involving $\mathcal{RK}$-symmetry breaking and several regimes with distinct magnetic correlations.

Overall, our exact analytical and numerical analysis shows that both the formalism used to define observables and the choice of eigenstate crucially influence the apparent critical properties of non‑Hermitian spin chains. Which of the possible phase diagrams is physically realized ultimately depends on the preparation protocol of the system in an (hypothetical) experiment, highlighting the need to carefully connect theoretical analyses of non‑Hermitian Hamiltonians to realistic measurement schemes.

\section*{Acknowledgments}
We would like to thank Apostolos Siskou for very useful discussion. In addition, we thank Bao-Ming Xu for providing the data of Ref.~\cite{Wang2025}.

\appendix
\section{Complex-$\lambda$ model: Subleading oscillations in the LL phase}\label{oscis}

\begin{figure}[tbp] 
	\begin{center} 
		\includegraphics[width=0.8\linewidth]{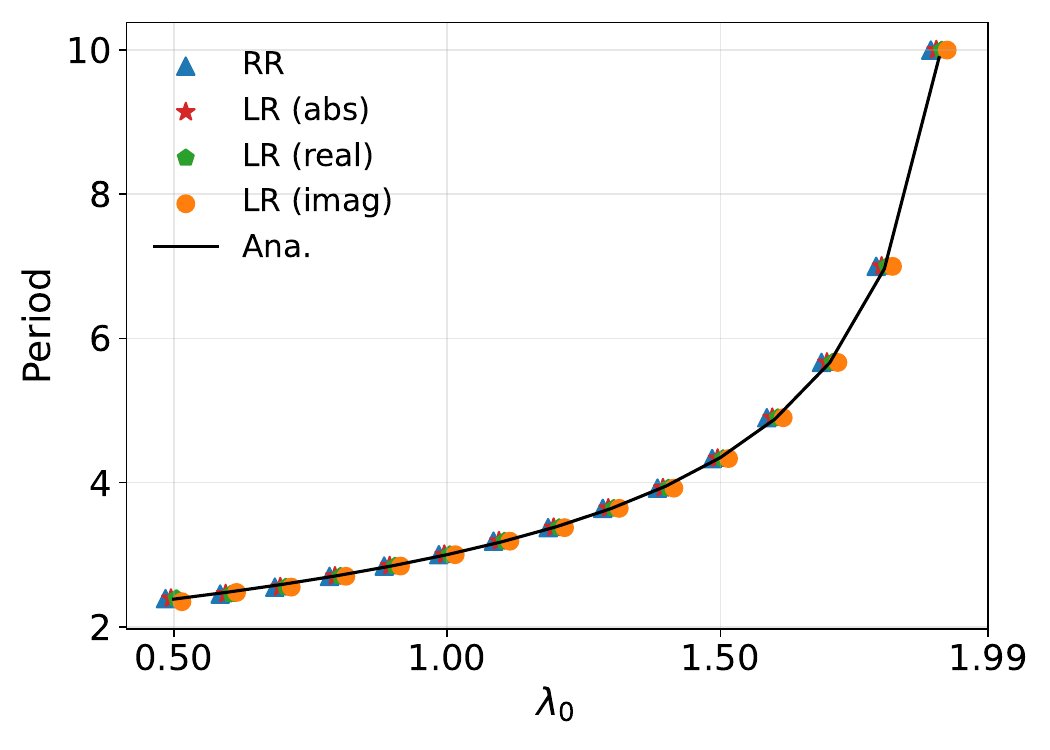} 
	\end{center}
	\vspace{-14pt}  
	\caption{The oscillation periods extracted from the numerical results of the RR and LR correlation functions $C_r^x$ for different $\lambda_0$, with $\lambda=\lambda_0\exp(\mi\pi/3)$ at $\gamma=0.25$.
	For clarity, the numerical data are horizontally shifted from $\lambda_0$ by -0.015, -0.005, 0.005, and 0.015.}        
	\label{per_LL}
\end{figure}

As already mentioned in the main part the right panels of Figs.~\ref{com_x_rr_lh_ga} and \ref{com_x_lr_lh_ga} reveal that, although the correlation functions decay monotonically at long distances, their log-log derivative exhibit oscillatory behavior as a function of distance, indicating that the oscillatory contributions arise as the subleading corrections to the leading power-law decay. As shown in Fig.~\ref{com_x_rr_lh_ga} (a2) and (b2), the oscillation period in the RR correlation functions depends on $\lambda$ but is independent of $\gamma$. From Fig.~\ref{com_x_lr_lh_ga}, it can be seen that the oscillation periods for the absolute value, real part, and imaginary part of the LR correlation functions are identical to each other. In Fig.~\ref{per_LL}, we plot the oscillation periods extracted from the RR and LR numerical data. Since the LR expectation values can be viewed as an analytic continuation of the corresponding Hermitian results, the oscillation period can be directly inferred from the Hermitian case as
\begin{equation}
T_{\text{LL}}=\pi/\arccos[\R(\lambda)].
\end{equation}
The periods obtained from both the RR and LR correlation functions are all consistent with this theoretical value.

\section{Fisher-Hartwig theorem}
\label{ll_ex}

The Fisher–Hartwig (FH) theorem characterizes the asymptotic behavior of Toeplitz determinants $D_L(\phi)$ as the matrix dimension $L\rightarrow \infty$ when the generating function $\phi(\theta)$ contains singularities on the unit circle. The Toeplitz determinant $D_L(\phi)$ is defined as the determinant of the $L \times L$ matrix
\begin{equation}
D_L(\phi) = \det(\phi_{i-j})_{0 \leq i,j \leq L-1}\label{adet}
\end{equation}
where the matrix elements are given by the Fourier coefficients of the generating function $\phi(\theta)$
\begin{equation}
\phi_l = \frac{1}{2\pi} \int_{0}^{2\pi} \phi(\theta) e^{-\mi l\theta} \dd \theta, \quad l \in \mathbb{Z}.\label{fh_f}
\end{equation}
The nonanalytic structure of $\phi(\theta)$ determines the asymptotic behavior of $D_L(\phi)$. 
In the FH framework, a generating function with $R$ singularities located at $\theta_r$ on the unit circle can be written as
\begin{equation}
\phi(\theta) = \psi(\theta) \prod_{r=1}^{R} [2 - 2\cos(\theta - \theta_r)]^{\alpha_r} e^{-\mi\beta_r(\pi - \theta + \theta_r)},\label{fhs}
\end{equation}
where $\psi(\theta)$ is a smooth non-vanishing function with zero winding number, and $\R(\alpha_r)>-1/2$ insures integrability. The parameters $\alpha_r$ and $\beta_r$ characterize root-type (zeros/poles) and jump (phase discontinuity) singularities, respectively. For $L \to \infty$, assuming $\max_{j,k} |\R(\beta_j)-\R(\beta_k)|<1$ and $\alpha_j\pm\beta_j\neq-1,-2,\cdots$, the determinant in Eq.~\eqref{adet} follows the scaling law
\begin{equation}
D_L(\phi(\theta)) \sim (\mathcal{F}[\psi])^L L^{\sum_{r=1}^{R} (\alpha_r^2 - \beta_r^2)} \mathcal{E}[\psi, \alpha_r, \beta_r, \theta_r],\label{fhd}
\end{equation}
where the power-law exponent $\sum_r (\alpha_r^2- \beta_r^2)$ is determined by all the  singrities. For a generating function of the form $\phi(\theta)e^{\mi \theta}$,  based on Eq.~\eqref{fhd}, the determinant scales as
\begin{equation}
D_L(\phi(\theta)e^{\mi \theta}) \sim D_L(\phi(\theta))\sum_{r=0}^{R}n^{-2\beta_r-1}\frac{\Gamma(1+\alpha_j+\beta_j)}{\Gamma(\alpha_j-\beta_j)}.\label{fhd_m}
\end{equation}
If $\max_{j,k} |\R(\beta_j)-\R(\beta_k)|\geq1$, $\beta_j$ need to be adjusted by adding or subtracting integers 
$n_j$ under the global constraint $\sum_j n_j=0$. The leading-order contribution to the Toeplitz determinant is then determined by the FH-representation that minimizes the sum $\sum_j[\R(\beta_j)+n_j]^2$.

\subsection{Complex-$\lambda$ model}

\subsubsection{RR and LR exponents in the LL phase}\label{fh_com1}

For the LR expectation value, the behavior of correlation functions is determined by the scalar Toeplitz matrix in Eq.~\eqref{det} with elements $\tilde{M}_{r}=\langle B_0 A_{r+1}\rangle$. To apply the FH theorem, we rewrite $\tilde{M}_{r} = \phi_{-r}$ in the form of a standard Fourier transformation \eqref{fh_f}
\begin{equation}
\phi_r=-\frac{1}{2\pi} \int_{0}^{2\pi} \frac{\lambda-\cos k+\mi \sin k}{\sqrt{(\lambda - \cos k)^2+(\gamma \sin k)^2}} e^{\mi k} e^{-\mi rk} \dd k.
\end{equation}
Thus, the corresponding generating function reads 
\begin{equation}
\phi(k)=-\frac{\lambda-\cos k+\mi \sin k}{\sqrt{(\lambda - \cos k)^2+(\gamma \sin k)^2}} e^{\mi k}=-e^{\mi \theta}e^{\mi k},\label{sx}
\end{equation}
where the angle $\theta_k$ is defined by $\tan(\theta_k)=\gamma \sin k/(\lambda - \cos k)$. Note that for $\gamma > 0$, the range of the real part of $\theta_k$ is restricted to $\R(\theta_k) \in [0,\pi]$ for $k\in[0,\pi]$, and $\R(\theta_k) \in [-\pi,0]$ for $k\in[-\pi,0]$. Thus $\theta$ jumps from $-\pi$ to $\pi$ at $k=0$, corresponding to $\beta_0=-1$. When $\R(\lambda)<1$, the denominator of $\tan(\theta_k)$ changes sign at $\pm k_\mathrm{s}=\pm\arccos [\R(\lambda)]$, leading to a discontinuity in $\R(\theta_{\pm k_\mathrm{s}})$. Specifically, $\theta_{k_\mathrm{s}} (\theta_{-k_\mathrm{s}})$ jumps from $\pi(0)$ to $0 (-\pi)$ as $k$ crosses $k_\mathrm{s} (-k_\mathrm{s})$. Therefore, the generating function contains total three FH jump singularities.  According to Eq.~\eqref{fhs}, the corresponding jump parameters satisfy $\beta_{k_\mathrm{s}}=\beta_{-k_\mathrm{s}}=1/2$. Given that $\max_{j,k} |\R(\beta_j)-\R(\beta_k)|\geq1$, the updated $\beta$'s are $\beta_0=0,\beta_{k_\mathrm{s}}=-\beta_{-k_\mathrm{s}}=\pm1/2$. Using Eq.~\eqref{fhd_m}, the total power-law exponent is \begin{equation}
-\beta_0^2-\beta_{k_\mathrm{s}}^2-\beta_{-k_\mathrm{s}}^2=-1/2.
\end{equation}

For the RR expectation value, the correlation functions are determined by the block Toeplitz matrix in Eq.~\eqref{pf} with elements Eq.~\eqref{M}. From the discussion of the LR expectation value, we know that the generating function contains only pure jump singularities. Therefore, the block Toeplitz problem can be reduced to a scalar one \cite{Basor2025}. The singularity at $k=0$ makes no contribution, as the phase change is $2\pi$. Around the singularity $k_\mathrm{s}$ (the analysis at $-k_\mathrm{s}$ is analogous), the left and right limits of Eq.~\eqref{M} towards $k_\mathrm{s}$ are 
\begin{eqnarray}
M_r^{-}&=&\begin{bmatrix}
\cos\theta_0 & \sin\theta_0 \e^{\mi k}   \\
-\sin\theta_0 \e^{-\mi k} &\cos\theta_0
\end{bmatrix},\no\\
M_r^{+}&=&\begin{bmatrix}
\cos\theta_0 & -\sin\theta_0 \e^{\mi k}   \\
\sin\theta_0 \e^{-\mi k} &\cos\theta_0
\end{bmatrix},
\end{eqnarray}
where $\theta_0$ is given in Eq.~\eqref{theta0}. The power-law exponent is thus given by
$
-\sum_{r=1}^2\sum_{j=1}^2 (\beta_r^{j})^2,
$
where $\beta_r^{j}$ are the eigenvalues of the jump matrix at each singularity
$
\ln \left[(M_r^{+})^{-1}M_r^{-}\right]/(2\pi \mi).
$
The eigenvalues of the jump matrix above are $\pm\theta_0/\pi$. Taking into account the contribution from the singularity at $-k_\mathrm{s}$ as well as the relation $\text{Pf}^2 \ldots =\det \ldots$, the final result is given in Sect.~\ref{sec:whylog} of  the main text.

\subsubsection{RR and LR exponents at the FM-LL transition point}\label{fh_com2}

In Fig.~\ref{com_y_rr_lh_ga} and Fig.~\ref{com_y_lr_ll_ga}, the power-law decay exponents of the $y$-correlation functions for the RR and LR expectation values at the FM–LL phase transition point are shown to be $0$ and $3$, respectively. 

For the LR expectation value, the generating function in the $y$-direction is given by $\phi(k)=-e^{-\mi \theta} e^{\mi k}$. In this scenario, three singularities emerge on the unit circle: one located at $k=0$, and the other two $k=\pm k_\mathrm{s}$ determined by the conditions $\cos k_\mathrm{s} = \R(\lambda)$ and $\gamma \sin k_\mathrm{s} = \I(\lambda)$. When $\gamma>0$, the singularity at $k=0$ exhibits a phase jump of $-2\pi$, corresponding to $\beta_0=1$. At $-k_\mathrm{s}$, the phase jump is $\pi/2$ and the modulus of the generating function scales as $|k+k_\mathrm{s}|^{-1/2}$, yielding the FH parameters $\beta_{-k_\mathrm{s}}=-1/4$ and $\alpha_{-k_\mathrm{s}}=-1/4$. Similarly, at $k_\mathrm{s}$, the phase jump is also $\pi/2$, while the modulus scales as $|k-k_\mathrm{s}|^{1/2}$, leading to $\beta_{k_\mathrm{s}}=-1/4$ and $\alpha_{k_\mathrm{s}}=1/4$. As the initial $\beta$ parameters differ by more than $1$, a shift of the $\beta$ is required. The updated values are $\beta_0=0$, $\beta_{-k_\mathrm{s}}=-1/4$ and $\beta_{k_\mathrm{s}}=3/4$, respectively. Moreover, due to the presence of the factor $e^{\mi k}$ in the generating function, Eq.~\eqref{fhd_m} dictates that another contribution arises from the term that maximizes the exponent $-2\beta-1$. Given the divergence of $\Gamma(0)$ in the denominator of Eq.~\eqref{fhd_m}, the leading contribution arises from $\beta_{k_\mathrm{s}}=3/4$. Combining these results, the power-law decay exponent for the LR $y$-correlation function at the FM–LL transition point is
\begin{equation}
\sum_{j=1}^3(\alpha_j^2-\beta_j^2)-2\beta_{k_\mathrm{s}}-1=-3.
\end{equation}

For the RR expectation value, the elements of the Toeplitz matrix in Eq.~\eqref{M} remain finite at $\pm k_\mathrm{s}$. Consequently, the generating function is devoid of singularities at these points. At $k=0$, the phase change is exactly $2\pi$, which gives no contribution. Due to the absence of algebraic singularities on the unit circle, the correlation function does not exhibit power-law decay, or, in other words, the exponent is 0. 

\subsection{Imaginary‑$\gamma$ model}\label{fh_imag}

For the LR expectation value in the $\mathcal{RK}$-broken phase, the singular behavior of the generating function in Eq.~\eqref{det} varies across three distinct regimes: $|\lambda| < 1$, $|\lambda| = 1$, and $1 < |\lambda| < \lambda_c$. The singular structures are summarized in Table~\ref{tab:singularities}. By applying Eq.~\eqref{fhs}, the corresponding FH exponents $(\alpha, \beta)$ for each singularity are obtained, as detailed in Table~\ref{tab:fh_exponents}. Therefore, the overall scaling exponents for the correlation functions can be derived using Eq.~\eqref{fhd_m}.

For the RR expectation value, the general formulation of the FH theorem for generating function containing zero/pole-type singularities has not yet been established.

\begin{table}[tbp]
	\centering
	\begin{tabular}{ccc}
		\toprule\toprule
		Regime & Singularities ($k_\mathrm{s}$) & Behavior \\
		\midrule
		$|\lambda| < 1$ & $-k_2, -k_1, k_1, k_2$ & $1/\sqrt{-\delta}, -\delta/\sqrt{\delta}, -1/\sqrt{-\delta}, \delta/\sqrt{\delta}$ \\ 
		\midrule
		$|\lambda| = 1$ & $-k_2, 0, k_2$ & $1/\sqrt{-\delta}, -\delta/\sqrt{-\delta^2}, \delta/\sqrt{\delta}$ \\ 
		\midrule
		$1 < |\lambda| < \lambda_c$ & $-k_2, -k_1, k_1, k_2$ & $1/\sqrt{-\delta}, 1/\sqrt{\delta}, -\delta/\sqrt{-\delta}, \delta/\sqrt{\delta}$\\
		\bottomrule\bottomrule
	\end{tabular}
	\caption{Singularity behavior for the LR correlation functions in the $\mathcal{RK}$-broken phase. Here $\delta = k - k_\mathrm{s}$, where $k_\mathrm{s}$ denotes the positions of the singularities.}
	\label{tab:singularities}
\end{table}

\begin{table}[tbp]
	\centering
	\begin{tabular}{ccc}
		\toprule\toprule
		Singularity &  $\alpha$ &  $\beta$ \\ 
		\midrule
		$1/\sqrt{-\delta}$ & $-1/4$ & $1/4 \cdot a$ \\
		$-\delta/\sqrt{\delta}$ & $1/4$ & $-3/4 \cdot a$ \\
		$-1/\sqrt{-\delta}$ & $-1/4$ & $1/4 \cdot a$ \\
		$\delta/\sqrt{\delta}$ & $1/4$ & $1/4 \cdot a$ \\
		$1/\sqrt{\delta}$ & $-1/4$ & $-1/4 \cdot a$ \\
		$-\delta/\sqrt{-\delta}$ & $1/4$ & $-1/4 \cdot a$ \\
		$-\delta/\sqrt{-\delta^2}$ & 0 & $-1/2 \cdot a$ \\
		\bottomrule
		\bottomrule
	\end{tabular}
	\caption{The FH exponents $(\alpha, \beta)$ for singularities summarized in Table~\ref{tab:singularities}. Here, $a=-1$ and $a=1$ correspond to the $x$- and $y$-correlation functions, respectively.}
	\label{tab:fh_exponents}
\end{table}

\section{Complex-$\lambda$ model: Details on the correlation functions in the PM phase}\label{com_pfy_pm}

\begin{figure}[tbp] 
	\begin{center} 
		\includegraphics[width=1\linewidth]{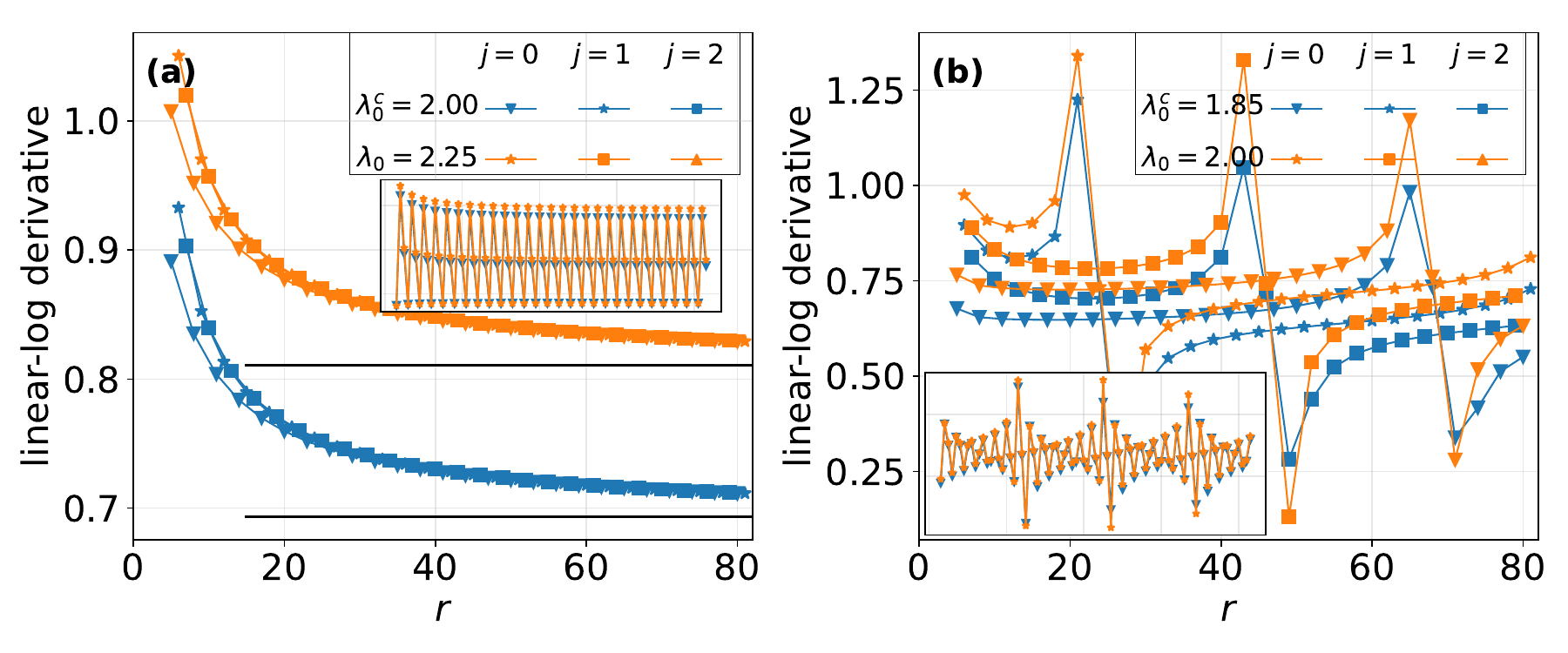} 
	\end{center}
	\caption{The linear-log derivative $-\dd\ln|C_r^x| / \dd r$ of Fig.~\ref{com_pfx_rr_pm} for different $\lambda_0$, with $\lambda=\lambda_0\exp(\mi\pi/3)$ (a) and $\lambda=\lambda_0\exp(\mi)$ (b), at $\gamma=1$ in the PM phase. The black lines are the analytical results Eq.~\eqref{pm_ex_ga1}. The numerical data are separated into three subsets according to $r \pmod 3$: $r = 3n+1, 3n+2$, and $3n$ (for $n \in \mathbb{N}$), labeled as $j=0, 1, 2$, respectively. The insets 
    show the raw, ungrouped exponents, which exhibit rapid oscillations around their respective mean values.}     
	\label{com_pfx_rr_pm_a}
\end{figure}

In the PM phase, the RR correlation functions exhibit oscillatory exponential decay as shown in Fig.~\ref{com_pfx_rr_pm}. For $\gamma^2=1$ and in the commensurate phase, the inverse correlation length can be numerically extracted by separating $|C_r^x|$ into several distinct sequences according to $r \pmod T$, where $T=\pi/\arg(\lambda)$ denotes the oscillation period. These subsequences are defined by $r=Tn+j$ with $n\in\mathbb{N}$ and labeled by $j=0, 1, \dots, T-1$. Within each sequence, the inverse correlation length is obtained from the linear-log derivative $-\dd \ln|C_r^x| / \dd r$. Fig.~\ref{com_pfx_rr_pm_a} shows the corresponding linear-log derivatives for the data in Fig.~\ref{com_pfx_rr_pm}. In the incommensurate phase with no well-defined period, extracting the correlation length becomes challenging.

\begin{figure}[tb]  
		\includegraphics[width=0.99\linewidth]{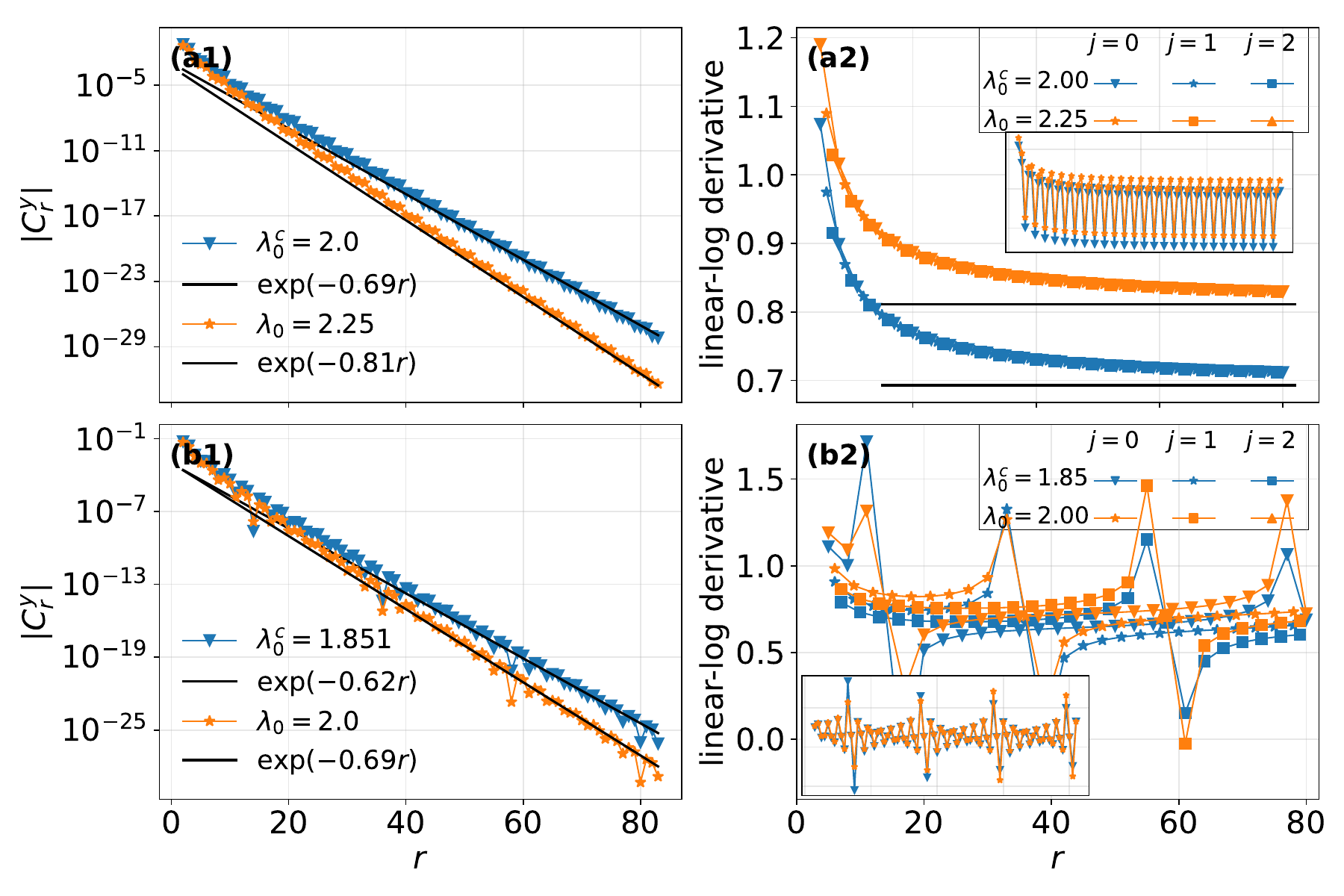} 
	\vspace{-14pt}  
	\caption{The $r$-dependence of the RR $y$-correlation functions $|C_r^y|$ for different $\lambda_0$, with $\lambda=\lambda_0\exp(\mi\pi/3)$ (upper row) and $\lambda=\lambda_0\exp(\mi)$ (lower row), at $\gamma=1$ in the PM phase, in comparison with Fig. \ref{com_pfx_rr_pm}. The black lines show analytical results for the inverse correlation length; see Eq. \eqref{pm_ex_ga1}.}    
	\label{com_pfy_rr_pm_a}
\end{figure}

\begin{figure}[tb]  
	\includegraphics[width=0.8\linewidth]{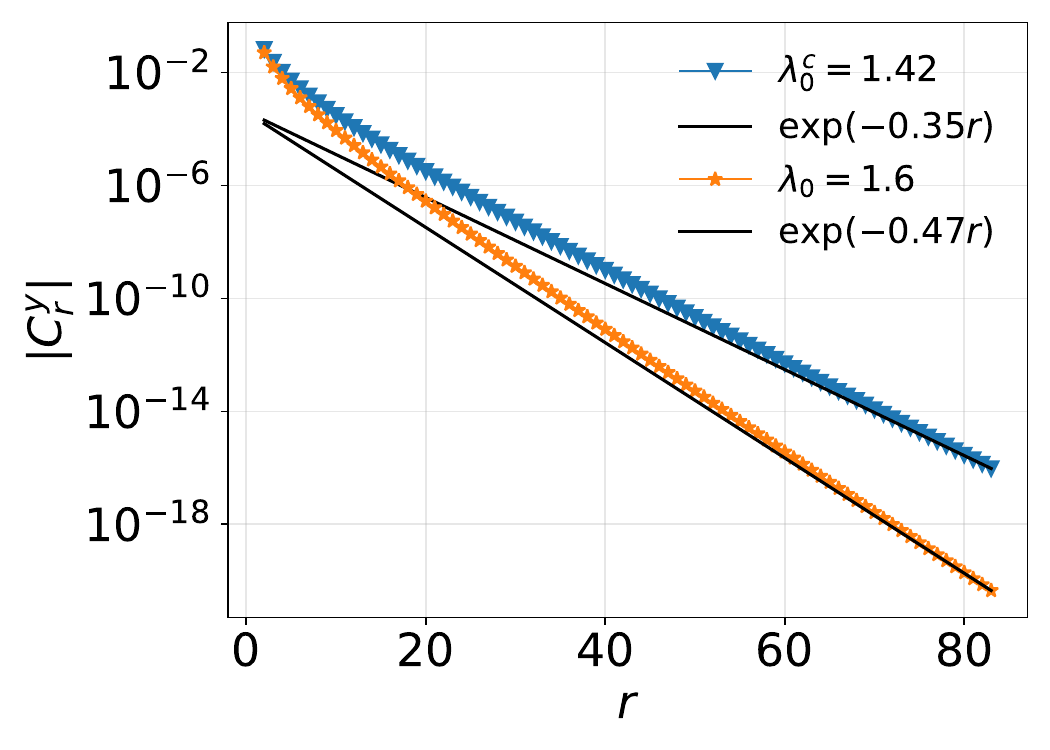} 
	\vspace{-14pt}  
	\caption{The $r$-dependence of the modulus of the LR $y$-correlation functions $|C_r^y|$ for different $\lambda_0$, with $\lambda=\lambda_0\exp(\mi\pi/4)$, at $\gamma=1$ in the PM phase, in comparison with Fig. \ref{com_pfx_lr_pm}. The black lines show analytical results for the inverse correlation length; see Eq. \eqref{pm_ex_ga1}.}    
	\label{com_pfy_lr_pm_a}
\end{figure}

The $y$-correlation function are shown in Figs.~\ref{com_pfy_rr_pm_a} and \ref{com_pfy_lr_pm_a}, in comparison with Figs.~\ref{com_pfx_rr_pm} and \ref{com_pfx_lr_pm}. They possess the identical inverse correlation length and oscillation structure as the $x$-correlation function.

\section{Imaginary-$\gamma$ model: Symmetry relations for $C_r^{x/y}$}
\label{app_sym}

Due to the presence of $\mathcal{RK}$ symmetry, the $x$-correlation function with imaginary energy branch $\pm\mi$ maps exactly to the $y$-direction with branch $\mp\mi$.
Since the Hamiltonian commutes with the $\mathcal{RK}$ operator, the right and left vacuum eigenstates of the two branches are connected by the $\mathcal{RK}$ action.
Given that $\mathcal{RK}$ is an antilinear operator, it follows that 
\begin{equation}
C_r^{x}(\pm \mi) = [C_r^{y}(\mp \mi)]^*.\label{s1}
\end{equation}
For the RR expectation value, which are purely real, the correlation functions for $x$ at branch $\pm \mi$ and $y$ at branch $\mp \mi$ are strictly identical. In contrast, for the LR expectation value, these correlation functions are related by complex conjugation. Similarly, a corresponding symmetry also exists for the anisotropic parameter $\gamma$
\begin{equation}
C_r^{x/y}(\gamma,\pm \mi) = [C_r^{x/y}(-\gamma,\mp \mi)]^*,\label{s2}
\end{equation}
which establishes a symmetry relation between $\pm\gamma$, $\pm \mi$, and $x/y$.

\begin{figure}[htbp] 
	\begin{center} 
		\includegraphics[width=1\linewidth]{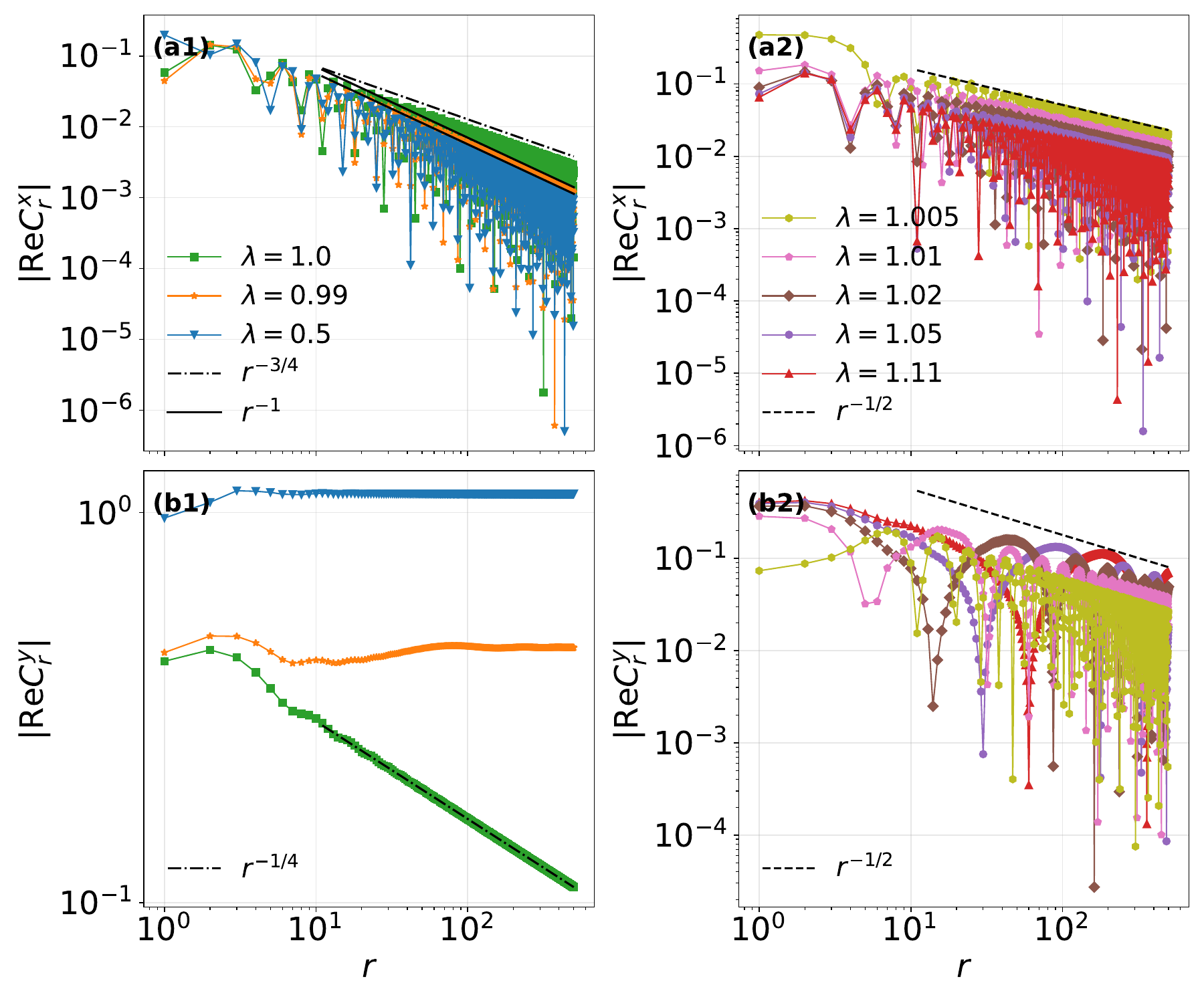} 
	\end{center}
	\vspace{-14pt}  
	\caption{The real parts of LR $x$-direction (upper row) and $y$-direction (lower row) correlation functions as a function of $r$ at $\gamma = 0.5$, in comparison with Fig. \ref{lr_lh}.}         
	\label{appre}
\end{figure}

\begin{figure}[htbp] 
	\begin{center} 
		\includegraphics[width=1\linewidth]{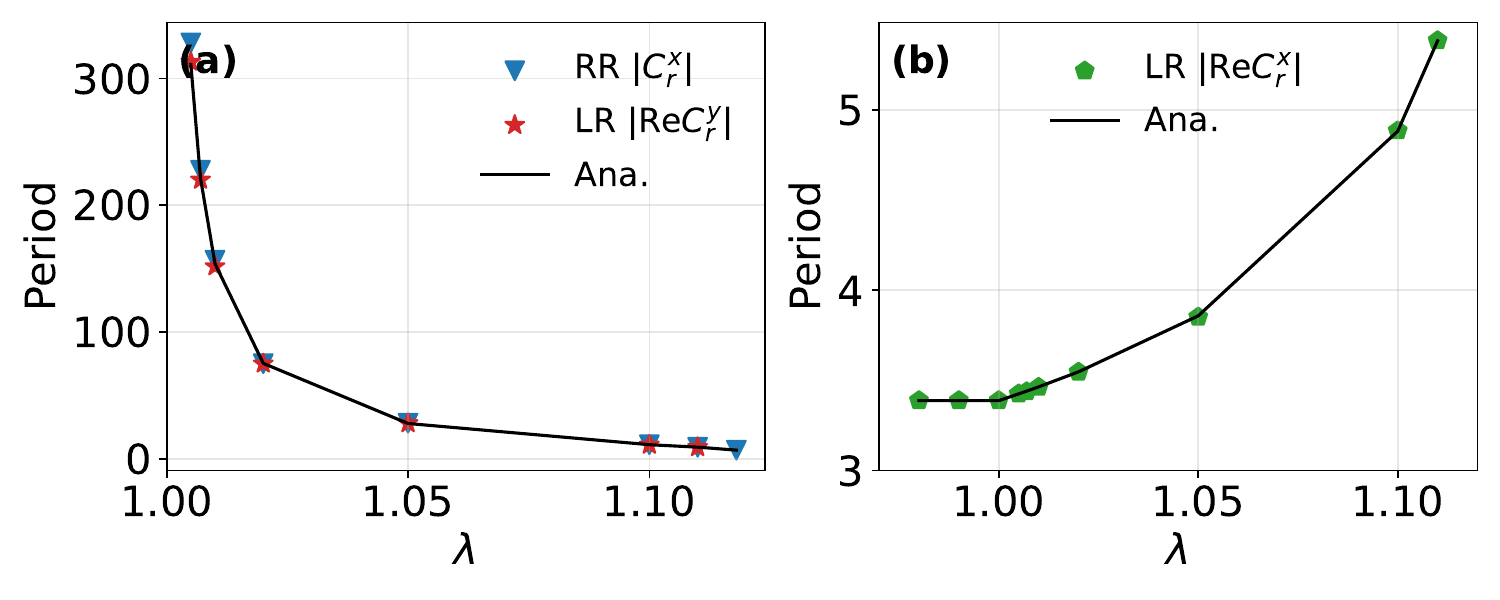} 
	\end{center}
	\vspace{-14pt}  
	\caption{The oscillation periods of RR and LR correlation functions at $\gamma=0.5$ for $0<|\lambda|\leq\lambda_c$. The analytical results are Eq.~\eqref{T}.}         
	\label{per}
\end{figure}

\section{Imaginary-$\gamma$ model: Oscillations in the $\mathcal{RK}$-broken phase}
\label{im-oscis}

As  illustrated in Fig.~\ref{rr_lh}, the RR correlation functions for $|\lambda| \leq 1$ exhibit only weak oscillations, with the primary oscillatory features appearing in the log-log derivative. This indicates that the oscillatory terms exist in the subleading order. In contrast, for $1 < |\lambda| \leq \lambda_c$, the oscillations become pronounced, and their frequency increases as the system approaches the $\mathcal{RK}$ symmetry-broken phase boundary, indicating that the oscillation terms emerge in the leading order. Regarding the LR correlations in Fig.~\ref{lr_lh}, the absolute values do not show significant oscillations throughout the $\mathcal{RK}$-broken phase. However, oscillations are present in their real and imaginary parts, as shown in Fig.~\ref{appre}. In particular, $|\R (C_r^{y})|$ only oscillates within $1 < |\lambda| < \lambda_c$, while $|\R (C_r^{x})|$ oscillates across the entire $\mathcal{RK}$-broken phase. Analytically, the oscillation periods can be derived from the poles of the energy spectrum, where the pole positions $k_0$ satisfy
\begin{equation}
\cos k_0^{\pm} =  \frac{\lambda \pm \gamma\sqrt{\gamma^2+1-\lambda^2}}{1+ \gamma^2}.\label{zeros}
\end{equation}
Both solutions remain real for $|\lambda|\leq\lambda_c$. The corresponding oscillation periods are given by 
\begin{equation}
T^{\pm} = \pi / k_0^{\pm}.\label{T}
\end{equation}
Here, $T^{+}$ characterizes the oscillation period of the RR correlation function $|C_r^{x}|$ and the LR counterpart $|\R(C_r^{y})|$ in the regime $1 < |\lambda| \leq \lambda_c$, while $T^{-}$ determines that of the LR correlation function $|\R(C_r^{x})|$ within the same parameter range. For $|\lambda| \leq 1$, the oscillation period of $|\text{Re}(C_r^{x})|$ remains constant, fixed by its value at $|\lambda| = 1$, which is given by $\pi/\arccos\big[(1-\gamma^2)/(1+\gamma^2)\big]$. At the transition point $\lambda=\lambda_c$, the two oscillation periods merge into an identical value, $T^{+}=T^{-}=\pi/\arctan(\gamma)$. A comparison between the oscillation periods extracted from the numerical data and the analytical results is shown in Fig.~\ref{per}.


\begin{thebibliography}{199}
\bibitem{footnote1} 
	These Lindblad operators also introduce overall imaginary constant shifts, $-\mi\I(\lambda)/2$ and $-\mi\gamma/2$, corresponding to a global decay rate (i.e., a shift on the zero-energy point) that does not affect physical observables after proper normalization.
\bibitem{footnote3} Note that in several papers \cite{Zhang2013a,Zhang2013b,Lakkaraju2021,Miao2024} the antilinear operator $\mathcal K$ is referred to as the time-reversal operator. This is misleading as the latter would also transform the spin-1/2 operators $\sigma_l^\alpha \to -\sigma_l^\alpha$.
\end{thebibliography}
\end{document}